\documentclass{aa}
\usepackage[varg]{txfonts}
\usepackage{graphicx}
\usepackage{natbib,twoopt}
\usepackage[breaklinks=true]{hyperref} 
\bibpunct{(}{)}{;}{a}{}{,} 
\makeatletter
 \newcommandtwoopt{\citeads}[3][][]{\href{https://ui.adsabs.harvard.edu/abs/#3/abstract}%
 {\def\hyper@linkstart##1##2{}%
 \let\hyper@linkend\@empty\citealp[#1][#2]{#3}}}
 \newcommandtwoopt{\citepads}[3][][]{\href{https://ui.adsabs.harvard.edu/abs/#3/abstract}%
 {\def\hyper@linkstart##1##2{}%
 \let\hyper@linkend\@empty\citep[#1][#2]{#3}}}
 \newcommandtwoopt{\citetads}[3][][]{\href{https://ui.adsabs.harvard.edu/abs/#3/abstract}%
 {\def\hyper@linkstart##1##2{}%
 \let\hyper@linkend\@empty\citet[#1][#2]{#3}}}
 \newcommandtwoopt{\citeyearads}[3][][]%
 {\href{https://ui.adsabs.harvard.edu/abs/#3/abstract}
 {\def\hyper@linkstart##1##2{}%
 \let\hyper@linkend\@empty\citeyear[#1][#2]{#3}}}
\makeatother

\begin{document}

   \title{The active centaur 2020~MK$_{4}$\thanks{Based on observations made with the 1m 
          Jacobus Kapteyn Telescope (JKT) at Observatorio del Roque de los Muchachos in 
          La Palma and the 82cm telescope of the Instituto de Astrofisica de Canarias 
          (IAC80) at Observatorio del Teide in Tenerife (Canary Islands, Spain).}
         }
   \author{C.~de~la~Fuente Marcos\inst{1}
            \and
           R.~de~la~Fuente Marcos\inst{2}
            \and
           J.~Licandro\inst{3,4}
            \and
           M.~Serra-Ricart\inst{3,4}
            \and
           S.~Martino\inst{5}
            \and
           J.~de~Le\'on\inst{3,4}
            \and
           F.~Chaudry\inst{3,4}
            \and
           M.~R. Alarc\'on\inst{3,4}
          }
   \authorrunning{de la Fuente Marcos et al.}
   \titlerunning{The active centaur 2020~MK$_{4}$}
   \offprints{C. de la Fuente Marcos, \email{nbplanet@ucm.es}}
   \institute{Universidad Complutense de Madrid,
              Ciudad Universitaria, E-28040 Madrid, Spain 
              \and
              AEGORA Research Group,
              Facultad de Ciencias Matem\'aticas,
              Universidad Complutense de Madrid,
              Ciudad Universitaria, E-28040 Madrid, Spain
              \and
              Instituto de Astrof\'{\i}sica de Canarias (IAC), 
              C/ V\'{\i}a L\'actea s/n, E-38205 La Laguna, Tenerife, Spain
              \and
              Departamento de Astrof\'{\i}sica, Universidad de La Laguna, 
              E-38206 La Laguna, Tenerife, Spain
              \and
              Departamento de Astronom\'{\i}a, Facultad de Ciencias, 
              Universidad de la Rep\'ublica, Igu\'a 4225, 11400, Montevideo, Uruguay 
             }
   \date{Received 7 August 2020 / Accepted 4 April 2021}

   \abstract
    {Centaurs go around the Sun between the orbits of Jupiter and Neptune. 
     Only a fraction of the known centaurs have been found to display 
     comet-like features. Comet 29P/Schwassmann-Wachmann~1 is the most 
     remarkable active centaur. It orbits the Sun just beyond Jupiter in a 
     nearly circular path. Only a handful of known objects follow similar 
     trajectories.}
    {We present photometric observations of 2020~MK$_{4}$, a recently found 
     centaur with an orbit not too different from that of 29P, and we perform 
     a preliminary exploration of its dynamical evolution.}
    {We analyzed broadband Cousins $R$ and Sloan $g'$, $r'$, and $i'$ images 
     of 2020~MK$_{4}$ acquired with the Jacobus Kapteyn Telescope and the 
     IAC80 telescope to search for cometary-like activity and to derive its 
     surface colors and size. Its orbital evolution was studied using direct 
     $N$-body simulations.}
    {Centaur 2020~MK$_{4}$ is neutral-gray in color and has a faint, compact 
     cometary-like coma. The values of its color indexes, 
     $(g'-r')=0.42\pm0.04$ and $(r'-i')=0.17\pm0.04$, are similar to the solar 
     ones. A lower limit for the absolute magnitude of the nucleus is 
     $H_{g}=11.30\pm0.03$~mag which, for an albedo in the range of 0.1--0.04, 
     gives an upper limit for its size in the interval (23,~37)~km. Its 
     orbital evolution is very chaotic and 2020~MK$_{4}$ may be ejected from 
     the Solar System during the next 200~kyr. Comet 29P experienced 
     relatively close flybys with 2020~MK$_{4}$ in the past, sometimes when 
     they were temporary Jovian satellites.}
    {Based on the analysis of visible CCD images of 2020~MK$_{4}$, we confirm
     the presence of a coma of material around a central nucleus. Its surface
     colors place this centaur among the most extreme members of the gray 
     group. Although the past, present, and future dynamical evolution of 
     2020~MK$_{4}$ resembles that of 29P, more data are required to confirm 
     or reject a possible connection between the two objects and perhaps 
     others.} 


   \maketitle

   \section{Introduction\label{intro}}
      Centaurs go around the Sun following unstable paths between the orbits of Jupiter and Neptune (see for example 
      \citealt{2007Icar..190..224D,2020ApJ...892L..38C}). While only a small fraction of the known centaurs have been found to 
      exhibit cometary activity in the form of moderately intense eruptions, one object has managed to remain continuously active 
      since its discovery nearly a century ago, experiencing semi-regular and comparatively very bright outbursts (see for 
      example \citealt{2009AJ....137.4296J,2012AJ....144...97G}). This object is 29P/Schwassmann-Wachmann~1 which orbits at a 
      distance between 5.7~AU and 6.3~AU from the Sun, beyond the region where water-ice sublimates efficiently (see for example 
      \citealt{2009AJ....137.4296J,2012AJ....144...97G,2020AJ....159..136W}). Its current orbit (see Table~\ref{elements}) is 
      rather unusual among those of minor bodies located beyond Jupiter as it has both low eccentricity, $e=0.0448$, and low 
      inclination, $i=9\fdg39$. On June 24, 2020, J.~Bulger, K.~Chambers, T.~Lowe, A.~Schultz, and M.~Willman observing with the 
      1.8-m Ritchey-Chretien telescope of the Pan-STARRS Project \citep{2004AAS...204.9701K} from Haleakala, discovered a close 
      orbital relative of 29P, 2020~MK$_{4}$, at an apparent magnitude $w$ of 19.8 \citep{2020MPEC....N...36D}. Its latest orbit 
      determination is shown in Table~\ref{elements}. 
%
%
      \begin{table*}
       \centering
       \fontsize{8}{12pt}\selectfont
       \tabcolsep 0.14truecm
       \caption{\label{elements}Values of the heliocentric Keplerian orbital elements and their respective 1$\sigma$ uncertainties 
                of comets 29P/Schwassmann-Wachmann~1, P/2008~CL94 (Lemmon), and P/2010~TO20 (LINEAR-Grauer), and centaur 
                2020~MK$_{4}$.
               }
       \begin{tabular}{lccccc}
        \hline
         Orbital parameter                                 &   & 29P/Schwassmann-Wachmann 1      & P/2008~CL94 (Lemmon) & P/2010~TO20 (LINEAR-Grauer) & 
                                                                 2020~MK$_{4}$         \\
        \hline
         Semimajor axis, $a$ (AU)                         & = &   5.99579143$\pm$0.00000004     &  6.1706$\pm$0.0002   &   5.6012$\pm$0.0003         & 
                                                                   6.15875$\pm$0.00007 \\
         Eccentricity, $e$                                 & = &   0.04437627$\pm$0.00000002     &  0.11941$\pm$0.00008 &   0.088533$\pm$0.000005     & 
                                                                   0.0220$\pm$0.0002   \\
         Inclination, $i$ (\degr)                          & = &   9.382089$\pm$0.000002         &  8.34819$\pm$0.00010 &   2.63909$\pm$0.00012       & 
                                                                   6.66716$\pm$0.00010 \\
         Longitude of the ascending node, $\Omega$ (\degr) & = & 312.595345$\pm$0.000011         & 33.4617$\pm$0.0004   &  43.9656$\pm$0.0008         & 
                                                                   2.344$\pm$0.002     \\
         Argument of perihelion, $\omega$ (\degr)          & = &  50.20600$\pm$0.00002           & 82.05$\pm$0.02       & 251.89$\pm$0.03             &
                                                                 176.2$\pm$0.2         \\
         Mean anomaly, $M$ (\degr)                         & = & 176.60347$\pm$0.00002           & 54.801$\pm$0.011     &  82.78$\pm$0.03             &
                                                                 112.7$\pm$0.2         \\
         Perihelion distance, $q$ (AU)                     & = &   5.72972056$\pm$0.00000013     &  5.4337$\pm$0.0004   &   5.1053$\pm$0.0002         &
                                                                   6.0230$\pm$0.0010   \\
         Aphelion distance, $Q$ (AU)                       & = &   6.26186229$\pm$0.00000004     &  6.9074$\pm$0.0002   &   6.0971$\pm$0.0003         &
                                                                   6.29447$\pm$0.00007 \\
         Absolute magnitude, $H$ (mag)                     & = &   8.6$\pm$1.0                   &  8.5$\pm$0.4         &   5.9$\pm$0.4               &
                                                                  11.4$\pm$0.6       \\
        \hline
       \end{tabular}
       \tablefoot{The orbit determination of comet 29P/Schwassmann-Wachmann~1 was computed by S.~Naidu, it is referred to 
                  as epoch JD 2455844.5 (2011-Oct-10.0) TDB (Barycentric Dynamical Time, J2000.0 ecliptic and equinox), and it is 
                  based on 33010 observations with a data-arc span of 8729 days (solution date, 2021-Feb-02 23:49:37 PST). The 
                  orbit determination of comet P/2008~CL94 (Lemmon) is referred to as epoch JD 2454769.5 (2008-Oct-30.0) TDB and it 
                  is based on 61 observations with a data-arc span of 491 days (solution date, 2020-Nov-11 10:56:31 PST). The 
                  orbit determination of comet P/2010~TO20 (LINEAR-Grauer) is referred to as epoch JD 2455832.5 (2011-Sep-28.0) TDB 
                  and it is based on 58 observations with a data-arc span of 776 days (solution date, 2020-Nov-11 10:56:01 PST).
                  The orbit determination of 2020~MK$_{4}$ is referred to as epoch JD 2459000.5 (2020-May-31.0) TDB and it is based 
                  on 108 observations with a data-arc span of 147 days (solution date, 2020-Dec-28 03:38:37 PST). Source: JPL's 
                  SBDB.
                 }
      \end{table*}
%
%

      The size and shape of the orbit of 2020~MK$_{4}$ are not too different from those of the orbit of 29P. Prior to the 
      discovery of 2020~MK$_{4}$, the two closest orbital relatives of 29P (see Table~\ref{elements}) were the comets P/2008~CL94 
      (Lemmon) with a semimajor axis, $a=6.171$~AU (29P has 5.9930~AU), $e=0.1194$, and $i=8\fdg35$ \citep{2009MPEC....F...28S,
      2016Icar..271..314K,2019AJ....157..225W} and P/2010~TO20 (LINEAR-Grauer) with $a=5.6006$~AU, $e=0.0887$, and $i=2\fdg64$ 
      \citep{2011CBET.2867....1G,2011CBET.2867....2S,2011MPEC....U...41P,2011IAUC.9235....1G,2013SoSyR..47..189E,
      2013MNRAS.428.1818L}. On the other hand, the announcement 
      MPEC\footnote{\url{https://www.minorplanetcenter.net/mpec/K20/K20N36.html}} of 2020~MK$_{4}$ also showed a significant 
      increase in brightness over the course of nearly a month which could be consistent with that of an active centaur in an 
      ourburst \citep{2020MPEC....N...36D}.\footnote{An extensive search for precovery images of 2020~MK$_{4}$ carried out by 
      S. Deen confirmed its absence down to $z<22$~mag around its expected location near perihelion back in 2016
      (\url{https://groups.io/g/mpml/message/35875}).} These two properties, an orbit similar to that of 29P and a possible rapid 
      brightness increase, led us to investigate further. In this work, we study the nature (asteroidal versus cometary) of 
      2020~MK$_{4}$ using photometry and perform a preliminary exploration of its past, present, and future dynamical evolution. 
      In Sect.~\ref{obs}, we describe the observations acquired and in Sect.~\ref{results}, we present the results of our 
      analysis. In Sect.~\ref{dynevo}, we explore the dynamical evolution of 2020~MK$_{4}$ and compare it with that of 29P and 
      related objects. Our findings are discussed in Sect.~\ref{discussion}. Finally, our conclusions are summarized in 
      Sect.~\ref{conclusions}.

   \section{Observations\label{obs}}
      We obtained CCD images of 2020~MK$_{4}$ with the 0.82~m IAC80 telescope at Teide Observatory on July 16 and July 24, 2020 
      and also with the 1.0~m JKT\footnote{\url{http://www.ing.iac.es/astronomy/telescopes/jkt/}} telescope operated by the 
      Southeastern Association for Research in Astronomy (SARA, \citealt{2017PASP..129a5002K}) on July 17, 2020. We used CAMELOT-2 
      (in Spanish, ``CAmara MEjorada Ligera del Observatorio del 
      Teide"-2)\footnote{\url{http://research.iac.es/OOCC/iac-managed-telescopes/iac80/camelot2-2/}} with the 
      IAC80\footnote{\url{http://research.iac.es/OOCC/iac-managed-telescopes/iac80/}} telescope, a camera with an e2V 231-84 
      4K$\times$4K pixels CCD, a 0.336~arcsec~pixel$^{-1}$ plate scale, and a 12{\farcm}3$\times$12{\farcm}3 effective 
      field of view. With the JKT, we used a 2K$\times$2K pixels ANDOR Ikon-L 2048 CCD camera 
      with a 0.34~arcsec~pixel$^{-1}$ plate scale and a 11{\farcm}6$\times$11{\farcm}6 field of view.

      On July 16 and with the IAC80, we obtained a series of images (exposure times of 120~s each) using the Cousins {\em R} filter 
      between 1:04 and 2:19 UT. On July 17 and with the JKT, we obtained a series of 93 images (exposure times of 90~s each) using 
      the Sloan {\em r'} filter between 0:13 and 2:37 UT. Finally, between July 23, 23:01 UT and July 24, 00:54 UT, with the 
      IAC80, we obtained a series of images (exposure times of 300~s each) using the Sloan {\em g', r',} and  {\em i'} filters (12 
      images in the {\em r'}, four in the {\em g'}, and four in the {\em i'}-band, doing four series of {\em r', g', r', i', r'} 
      images). We used sidereal tracking and the individual exposure time --- in particular the first two nights --- was selected 
      to ensure that the movement of the comet was smaller than the value of the seeing to avoid traces. Images were bias and 
      flat-field corrected (using sky flats). Observational circumstances are summarized in Table~\ref{observations}. On July 24, 
      we also observed the Landolt standard field star Mark A \citep{1992AJ....104..340L} to derive an absolute photometric 
      calibration. 
%
%
      \begin{table*}
       \centering
       \fontsize{8}{12pt}\selectfont
       \tabcolsep 0.15truecm
       \caption{\label{observations}Circumstances of observation of centaur 2020~MK$_{4}$.} 
       \begin{tabular}{lccccccccc}
        \hline
         Date       & Tel.  & UT-range        & X            & $r_h$ & $\Delta$ & $\alpha$ & $\theta_{\sun}$ & $\theta_{-V}$ & 
                      Filters          \\
                    &       & (h:m)           &              & (AU)  & (AU)     & (\degr)  & (\degr)         & (\degr)       &     
                                       \\
        \hline
         July 16    & IAC80 &  01:04 -- 02:19 & 1.80 -- 1.85 & 6.228 & 5.220    & 1.3      & 322.5           & 255.3         & 
                      {\em R}          \\
         July 17    & JKT   &  00:13 -- 02:37 & 1.80 -- 1.96 & 6.228 & 5.220    & 1.2      & 329.8           & 255.3         & 
                      {\em r'}         \\
         July 23/24 & IAC80 &  23.01 -- 00:54 & 1.80 -- 2.15 & 6.229 & 5.223    & 1.4      &  23.2           & 255.7         & 
                      {\em g', r', i'} \\
        \hline
       \end{tabular}
       \tablefoot{Information includes the date, airmass (X), heliocentric ($r_h$) and geocentric ($\Delta$) distances, phase angle 
                  ($\alpha$), position angle of the projected anti-Solar direction ($\theta_{\sun}$), and the position angle of 
                  the projected negative heliocentric velocity vector ($\theta_{-V}$). Orbital values have been taken from JPL's 
                  HORIZONS system and they are the averages within the indicated UT-range.}
      \end{table*}
%
%

   \section{Results\label{results}}
      \subsection{Cometary-like activity}
\begin{figure*}
        \centering
        \includegraphics[width=0.49\linewidth]{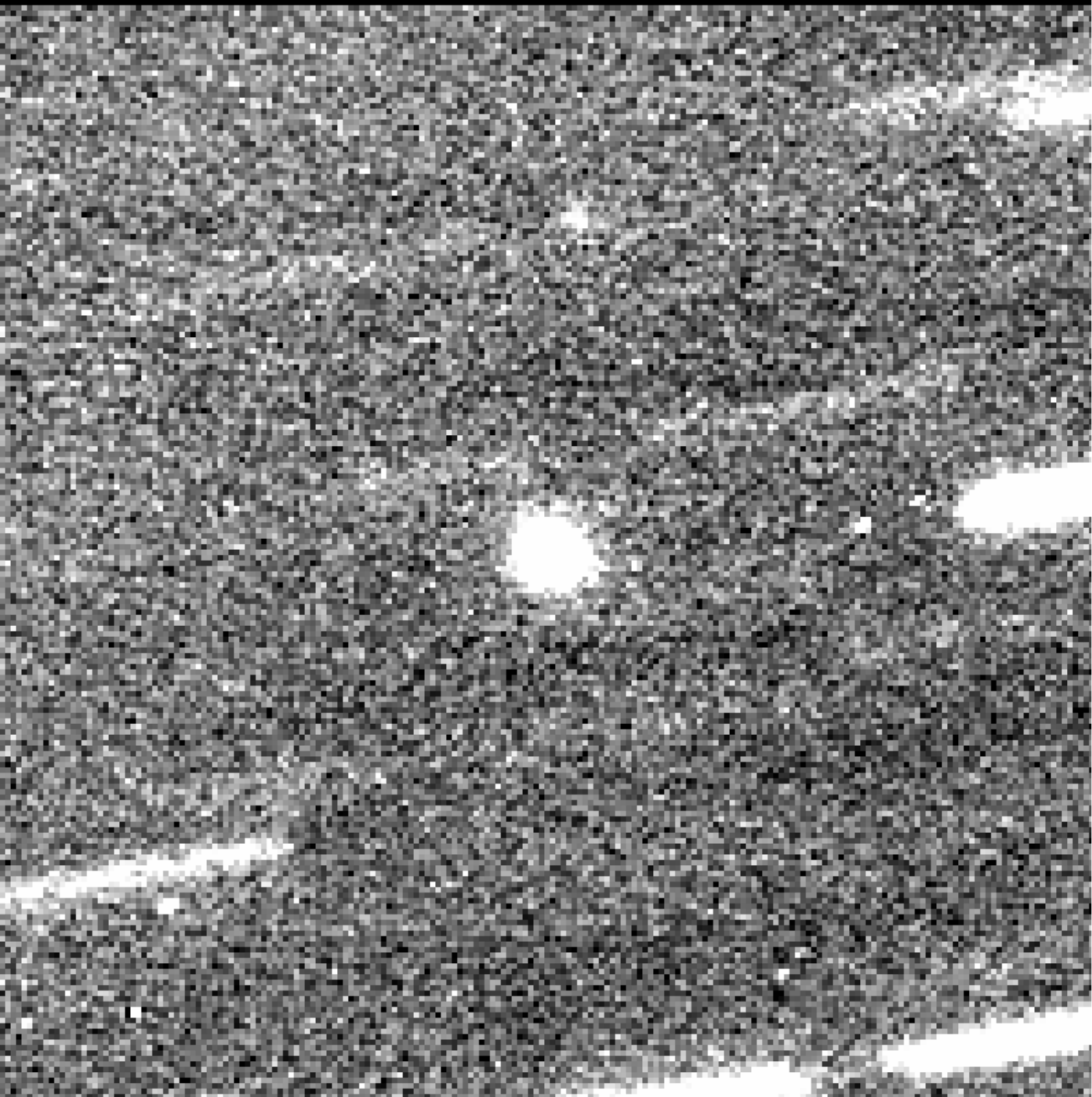}
        \includegraphics[width=0.49\linewidth]{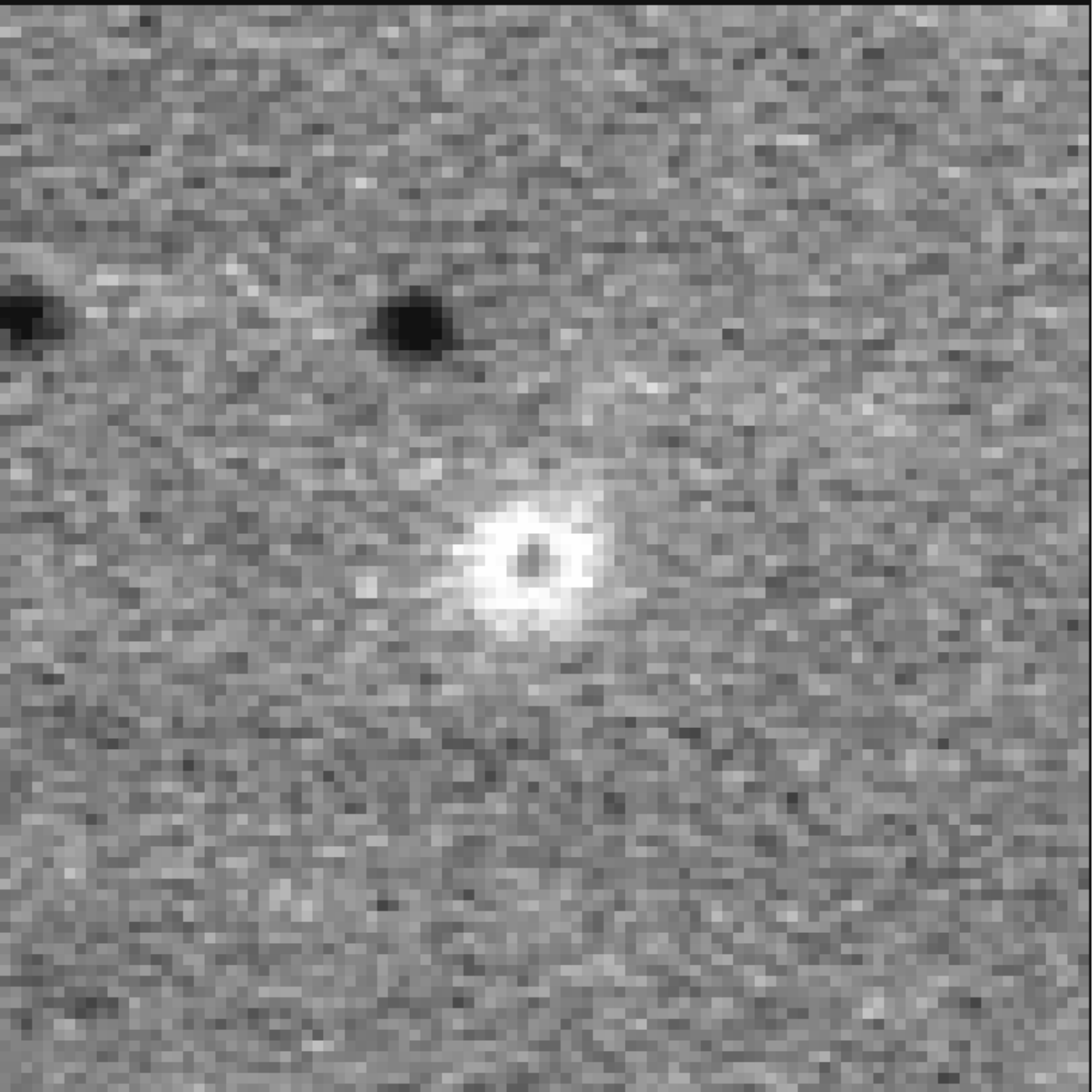}
        \caption{Observations of 2020~MK$_{4}$. {\em Left panel}: Image of 2020~MK$_{4}$ obtained on July 17, 2020 with the JKT 
                 telescope. This image is a combination of 93 images (exposure time of 90~s each), realigned so 2020~MK$_{4}$ was 
                 lined up. The field is 68\arcsec$\times$68\arcsec; north is up, and east is to the left. {\em Right panel}: Image 
                 combined on the comet minus the same combination of images centered on a bright star. Both images were sky 
                 subtracted and normalized to the peak before subtraction. It is important to notice that there is a circular 
                 residual compatible with the presence of a compact faint coma around 2020~MK$_{4}$. In this case, the field of 
                 view is 34\arcsec$\times$34\arcsec. The coma has a diameter of about 4\farcs1 or 15\,500~km at the geocentric 
                 distance indicated in Table~\ref{observations}.}
        \label{images}
     \end{figure*}

\begin{figure*}
        \centering
        \includegraphics[width=0.49\linewidth]{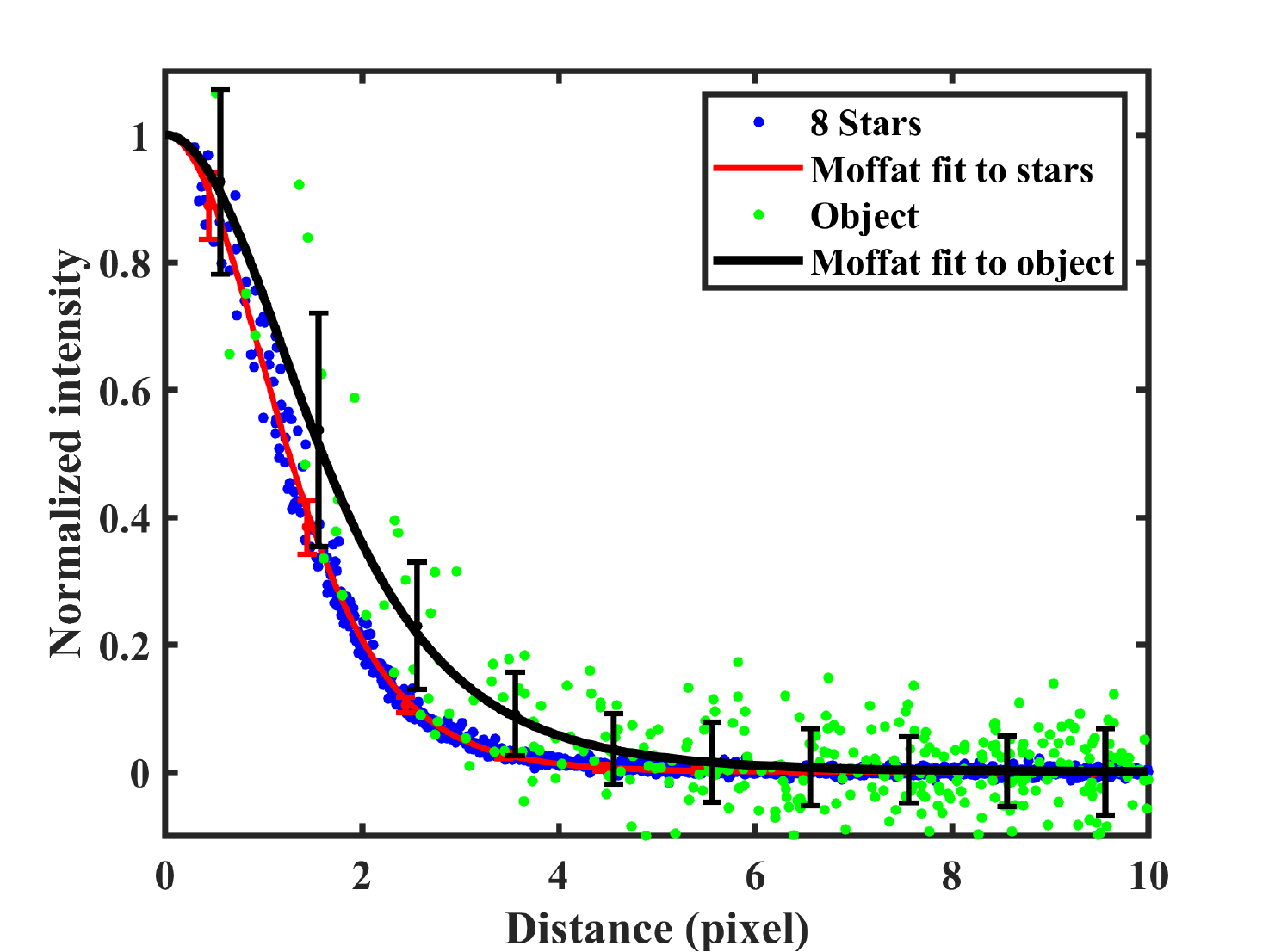}
        \includegraphics[width=0.49\linewidth]{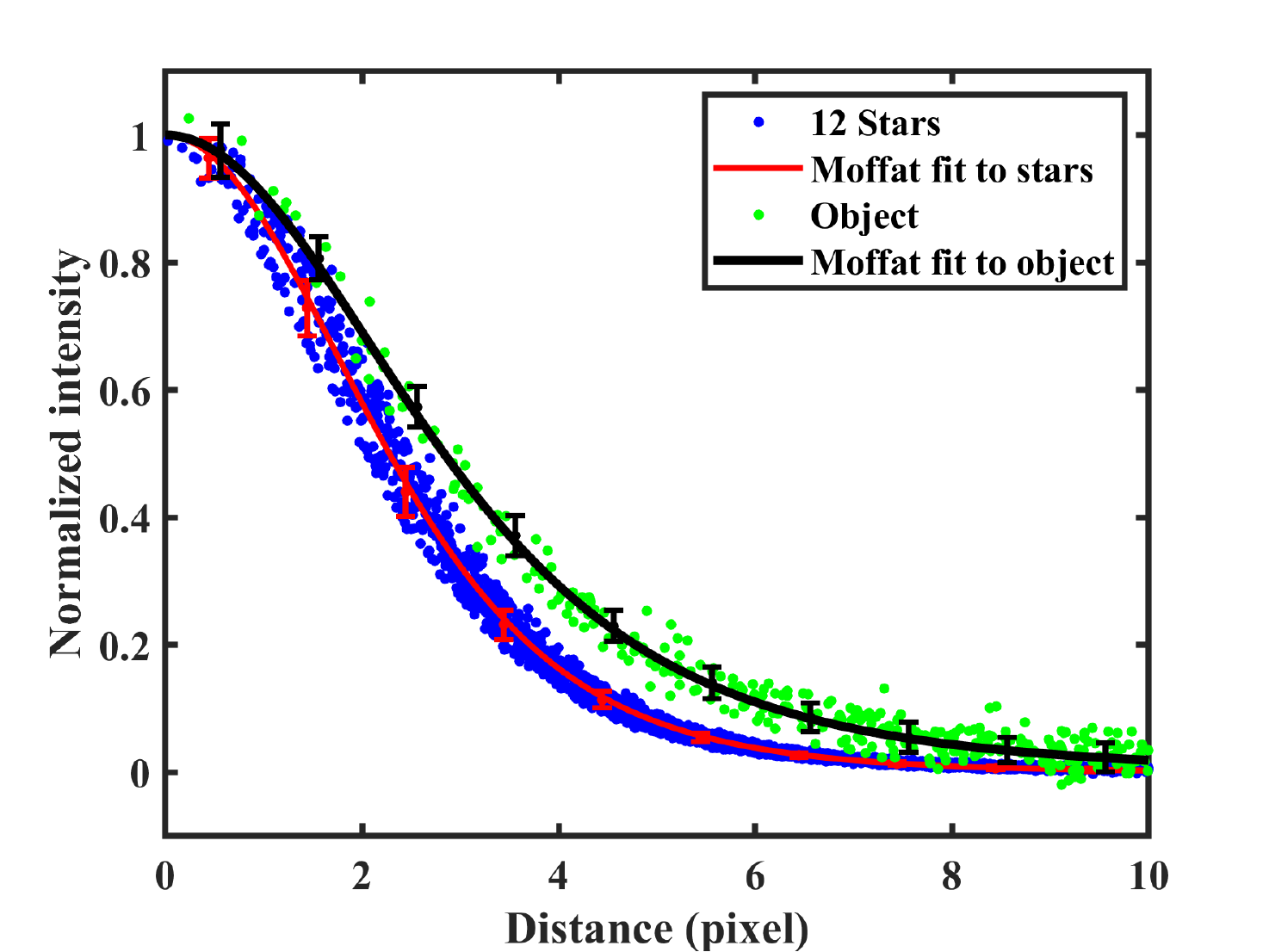}
        \caption{Profile analysis of the combined images of 2020~MK$_{4}$ obtained on July 16 (left panel) and July 17 (right 
                 panel). Each plot contains the values of the normalized intensity as a function of the distance from the centroid (in 
                 pixels) for the stars (blue dots), the Moffat fit for the stars (red line), values of normalized intensity for 
                 the object (green dots), and the Moffat fit for the object (black line). Red bars correspond to the stars while 
                 black bars correspond to the object, and they are placed in the middle of the interval; an artificial offset in the 
                 x-axis was introduced to avoid overlapping symbols. It is important to notice that the profile of 2020~MK$_{4}$ is significantly 
                 wider both nights, confirming that 2020~MK$_{4}$ shows clear signs of cometary-like activity.}
        \label{profiles}
     \end{figure*}

         During the analysis of the images of 2020~MK$_{4}$, we immediately noticed that the full width at half maximum (FWHM) of 
         the point spread function (PSF) of 2020~MK$_{4}$ was systematically wider than that of the stars during the first two 
         nights of observations, suggesting the presence of a faint, compact cometary-like coma. In order to determine if the 
         object was active, we made a direct comparison of its surface brightness profile with the profiles of the stars in the 
         field. We applied the following steps. First, we aligned the images so that the stars were all lined up and stacked onto 
         one another. A set of bright stars in the combined image was selected and, for these stars, the intensity versus distance 
         to centroid was extracted, normalized to the maximum, and combined to retrieve a high signal-to-noise ratio (S/N) stellar 
         profile. In a second step, all the images were realigned, so then 2020~MK$_{4}$ was lined up and then combined (see 
         Fig.~\ref{images}). The profile of the object was subsequently extracted from the combined image and normalized. Finally, 
         the resulting stellar profile and that of the object were fitted with a Moffat function and plotted together to compare 
         them. In Fig.~\ref{profiles}, we show the analysis of the profile of the combined images obtained on July 16 and 17. During 
         both nights, the brightness profile of 2020~MK$_{4}$ was significantly wider, indicating that 2020~MK$_{4}$ had a coma.  
%
%
%
%
%
%
%
%

         We also used the image with a higher S/N obtained on July 17 to check if the difference between the object and star profile 
         could be due to the proper motion of the object. In an attempt to see the coma, we subtracted the image of a bright star 
         from the combined image in which the stars were aligned from that of 2020~MK$_{4}$ in the combined image for which 
         2020~MK$_{4}$ was lined up. In order to do that, we subtracted the sky value and normalized the images to the brightness 
         peak of the object and star profile, respectively, carefully aligning the object and star, and we subtracted the star image. The 
         result of this procedure is shown in Fig.~\ref{images}. The resulting image presents an almost perfectly circular, but 
         doughnut-shaped residual. This corresponds to the expected shape of a compact coma and not to the effect of the proper 
         motion of the object that would produce residuals only in the direction of the motion. Therefore, we conclude that 
         2020~MK$_{4}$ was active at the time of the observations. 

      \subsection{Photometry and colors}
         In order to derive a limit for the absolute magnitude and obtain the colors of the object, we did aperture photometry of 
         the combined images for each night using standard tasks in the Image Reduction and Analysis Facility (IRAF).\footnote{IRAF 
         is distributed by the National Optical Astronomy Observatory, which is operated by the Association of Universities for 
         Research in Astronomy, Inc., under a cooperative agreement with the National Science Foundation.} We followed a procedure 
         similar to the one described in \citet{2019A&A...625A.133L}. We used an aperture diameter equivalent to the object's 
         FWHM. We obtained the absolute calibration using field stars with Sloan {\em g', r',} and  {\em i'} magnitudes determined in the 
         Pan-STARRS catalogue\footnote{\url{https://catalogs.mast.stsci.edu/panstarrs/}} and, on July 24, also using the flux 
         calibrated Landolt stars in the field of the star Mark~A and the transformation equations from \citet{2005AN....326..321B} 
         and \citet{2006AJ....132..989R} when needed. We obtained a magnitude of $r'=18.73\pm0.07$, $r'=18.88\pm0.03$, and 
         $r'=18.84\pm0.03$ on July 16, 17, and 24, respectively, and the colors $(g'-r')=0.42\pm0.04$ and $(r'-i')=0.17\pm0.04$ on 
         July 24. 

         Using Eq.~(1) from \citet{2019ApJ...886L..29J}, we derived a lower limit for the absolute magnitude of 2020~MK$_{4}$,  
         $H_{g}=11.30\pm0.03$~mag. Assuming a value of the visible geometric albedo between 0.1 and 0.04, this value of $H_{g}$ 
         provides an upper limit for the radius of the nucleus of the object, $R_{\rm N}$, between 23~km and 37~km. Using this 
         value of $H_g$ and assuming that this is the real absolute magnitude of the centaur, the apparent magnitude of 
         2020~MK$_{4}$ during June 2020 could have been $19.0$~mag$<r'<18.8$~mag. However, the observed brightness reported by 
         Pan-STARRS in \citet{2020MPEC....N...36D} shows that the object was 1 or 2 magnitudes fainter, with values around 21~mag 
         (in $w$) early in June and 19.9~mag at the end of the month, indicative of an activation around early June or perhaps 
         earlier than that. 

\begin{figure*}
        \centering
        \includegraphics[width=0.49\linewidth]{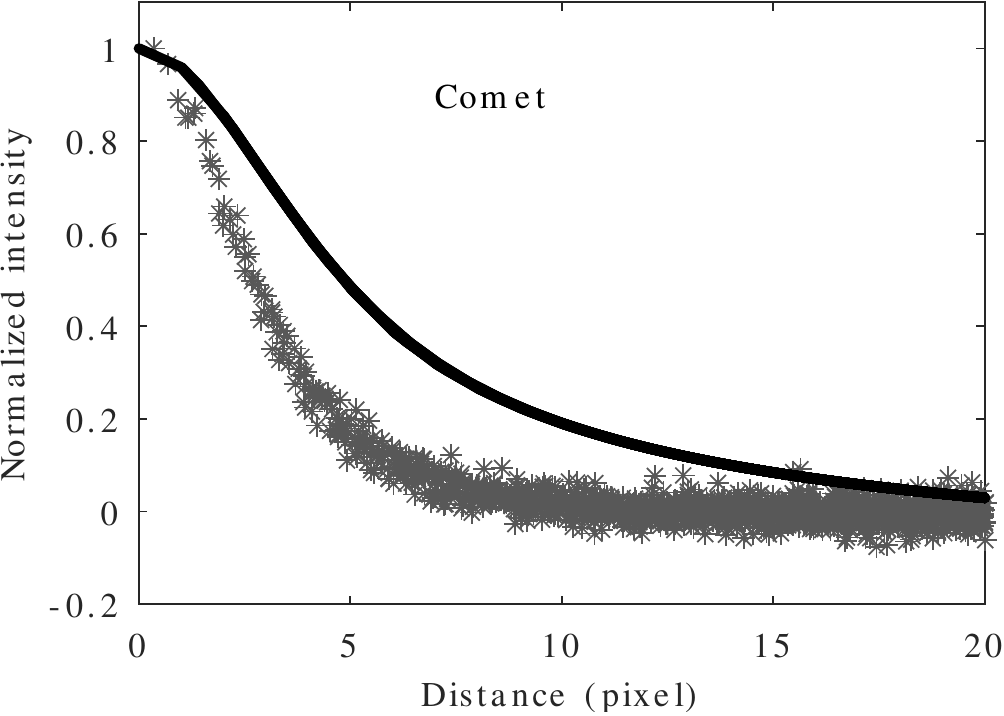}
        \includegraphics[width=0.49\linewidth]{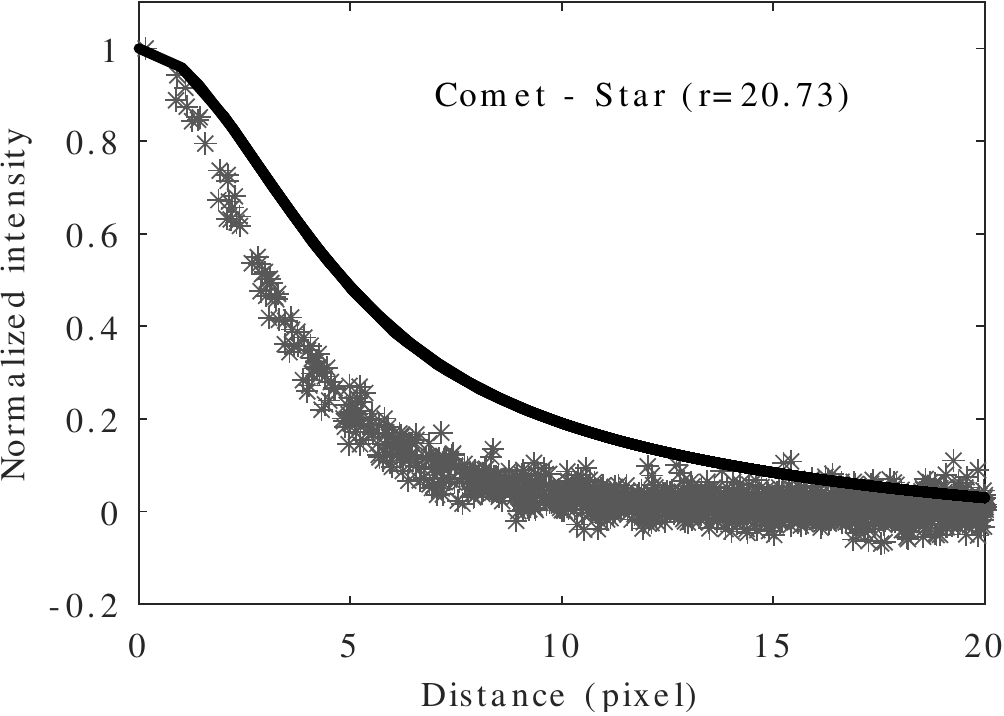}\\
        \includegraphics[width=0.49\linewidth]{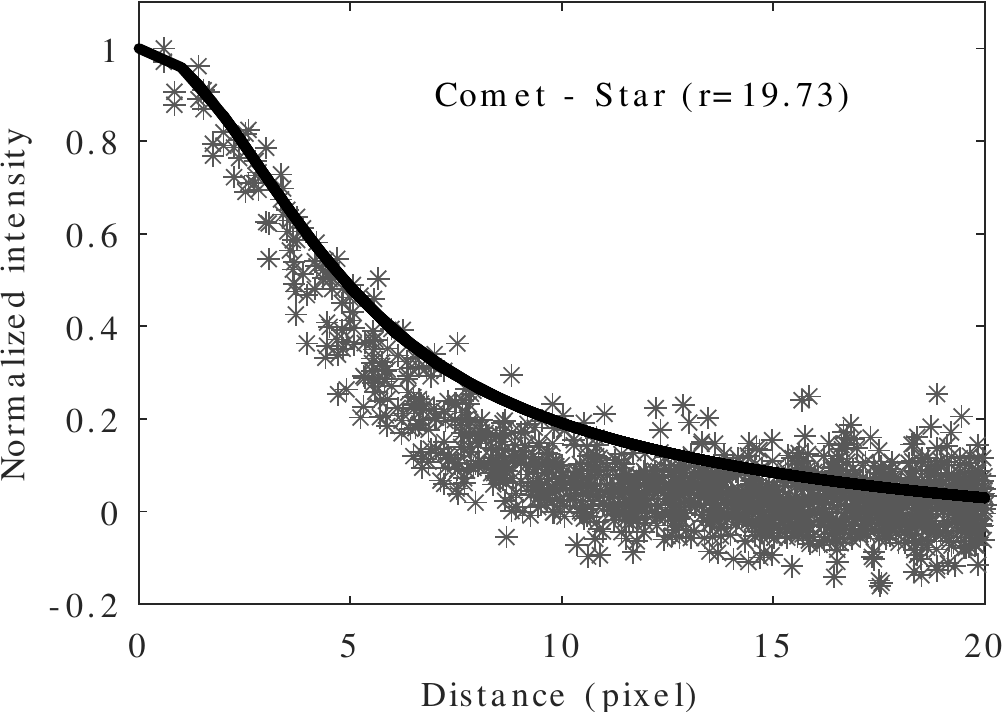}
        \includegraphics[width=0.49\linewidth]{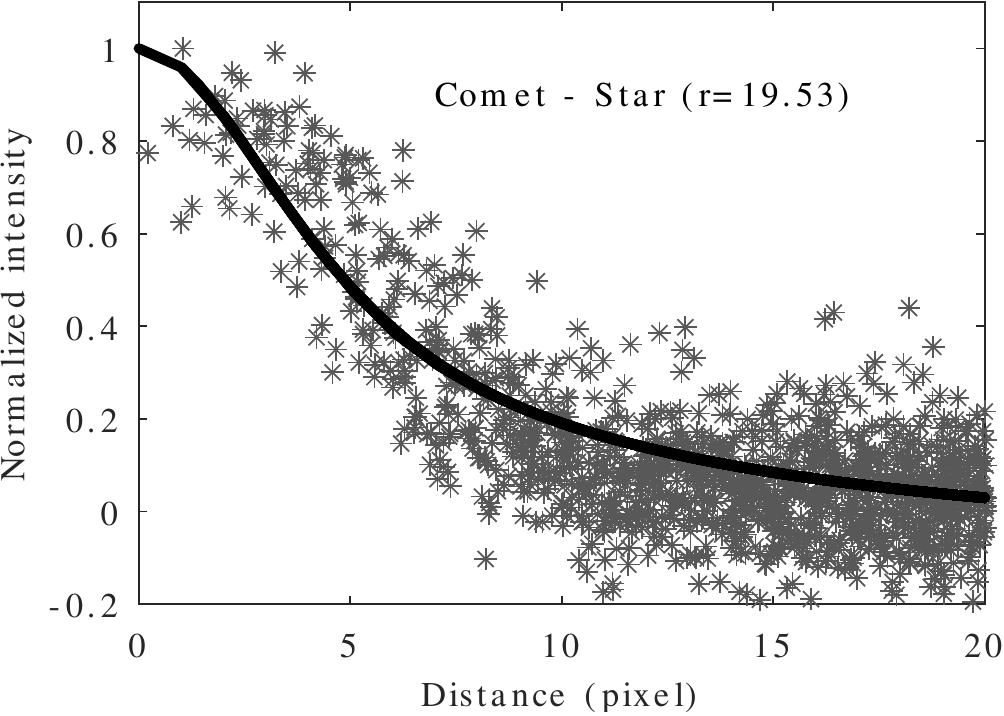}\\
        \caption{Evaluating the coma of 2020~MK$_{4}$. Normalized radial profile of the combined images of 2020~MK$_{4}$ obtained 
                 on July 17 minus a star profile of a different brightness (the assumed nucleus) together with the normalized 
                 radial profile of an  isotropic coma (thick, black curve, see text). {\em Upper-left panel}: The profile of the 
                 comet. {\em Upper-right panel}: The profile of the comet minus that of a nucleus (a star) of $r'$=20.73~mag. 
                 {\em Bottom-left panel}: The profile of the comet minus that of a nucleus of $r'$=19.73~mag. {\em Bottom-right 
                 panel}: The profile of the comet minus that of a nucleus of $r'$=19.53~mag. It is important to notice that the 
                 coma that remains once a nucleus of $r'$=19.53~mag has been subtracted resembles what an isotropic coma should 
                 look like.}
        \label{profiles2}
     \end{figure*}

         In order to make an approximate evaluation of the relative contribution of the nucleus to the total flux in the used 
         aperture, we subtracted a bright field star of known {\em r'} magnitude from the comet images as we did above. In this 
         case, we first scaled the star to $r'$=20.73, 19.73, and 19.53 magnitude (in other words, 2.0, 1.0, and 0.8 magnitudes 
         fainter than that of the comet). The normalized radial profiles of the resulting images are shown in Fig.~\ref{profiles2} 
         together with the profile of a simulated isotropic coma (thick, black curve). The profile of an isotropic coma was 
         generated using the {\em mkobject} task of IRAF, a $1/\rho$ profile for the coma, and a Moffat profile with the same 
         parameters obtained from the field stars to simulate the seeing. Figure~\ref{profiles2} suggests that the nuclear 
         magnitude is $\sim$0.8--1.0~mag fainter than the limit we obtained, that is  $H_g\sim12.30$~mag, and thus the nucleus 
         contributes about 66\% to 52\% of the total flux in the used aperture. If the nucleus is brighter than 19.53~mag, then a 
         ``hole" in the comet's profile should appear at the center; if it is fainter, then the coma profile should be much more 
         compact than the isotropic assumption. Our results suggest a value of $R_{\rm N}$ between 15~km and 23~km. 
%
%
%
%

         On the other hand, we noticed that the colors of 2020~MK$_{4}$ are similar to the solar ones\footnote{Solar colors are 
         $(g'-r')=0.44\pm0.02$, $(r'-i')=0.11\pm0.02,$ and $(i'-z')=0.02\pm0.03$ (see 
         \url{https://www.sdss.org/dr12/algorithms/ugrizvegasun/}).}. The colors of 2020~MK$_{4}$ correspond to those of the 
         members of the group of the gray centaurs (see \citealt{2003A&A...410L..29P,2003ApJ...599L..49T,2008ssbn.book..105T,
         2016AJ....152..210T}); in fact, it is perhaps one of the less red centaurs discovered thus far. It is well known that 
         centaur objects exhibit a peculiar physical property and that their visual colors divide the population into two distinct 
         groups: gray and red centaurs. The colors of 2020~MK$_{4}$ are also consistent with those observed in other active 
         centaurs; they all belong to the gray population, with the exception of (523676) 2013~UL$_{10}$, a red centaur 
         \citep{2018A&A...620A..93M}. \citet{2012A&A...539A.144M} showed that the different thermal reprocessing on the surface of 
         bodies of the red group on one side and the active and gray groups on the other is responsible for the observed 
         bimodality in the distribution of the surface colors of the centaurs; the color distribution of the gray centaurs is 
         similar to that of comet nuclei because gray centaurs likely had cometary activity. As we discussed above, the flux in 
         the used aperture is not just due to the brightness of the nucleus, but to the nucleus plus the dust coma. The coma can 
         contribute with $\sim$50\% of the total flux and this can also have some effect on the measured color. In any case and in 
         order to blue a red centaur so that it almost has a neutral color, the intrinsic color of the coma should be unusually 
         blue. However, it is clear that once its active nature has been confirmed, further observations with a higher S/N are 
         needed to understand the behavior of this centaur better.

   \section{Context and orbital evolution\label{dynevo}}
      Centaur 2020~MK$_{4}$ has been confirmed as active, but we still have a question regarding the similarity between its orbit 
      and that of comet 29P/Schwassmann-Wachmann~1 and related objects (see Table~\ref{elements}). The characterization of its 
      orbital context requires the study of the present-day orbital architecture of the sample of known objects that populate this 
      region of the orbital parameter space. Here, such a study is carried out by exploring the distributions of mutual nodal 
      distances and orientations in space. In order to analyze the results, we produced histograms using the Matplotlib library 
      \citep{2007CSE.....9...90H} with sets of bins computed using NumPy \citep{2011CSE....13b..22V,2020NumPy-Array} by applying 
      the Freedman and Diaconis rule \citep{Freedman1981}; kernel density estimations were carried out using the Python 
      library SciPy \citep{2020SciPy-NMeth}. 

      On the other hand, the assessment of the past, present, and future orbital evolution of 2020~MK$_{4}$ should be based on the 
      statistical analysis of results from a representative sample of $N$-body simulations. Here, such calculations were 
      carried out using a direct $N$-body code implemented by \citet{2003gnbs.book.....A} that is publicly available from the 
      website of the Institute of Astronomy of the University of 
      Cambridge.\footnote{\url{http://www.ast.cam.ac.uk/~sverre/web/pages/nbody.htm}} This software uses the Hermite integration 
      scheme described by \citet{1991ApJ...369..200M}. Results from this code compare well with those from 
      \citet{2011A&A...532A..89L} among others, as extensively discussed by \citet{2012MNRAS.427..728D}. 

      The initial conditions used in our calculations come from the orbit determination in Table~\ref{elements} which has been 
      released by Jet Propulsion Laboratory's Solar System Dynamics Group Small-Body Database (JPL's SSDG 
      SBDB).\footnote{\url{https://ssd.jpl.nasa.gov/sbdb.cgi}} Input data used in our orbital context analysis and in our 
      simulations were obtained from JPL's HORIZONS online solar system data and ephemeris computation service 
      \citep{2011jsrs.conf...87G,2015IAUGA..2256293G}.\footnote{\url{https://ssd.jpl.nasa.gov/?horizons}} Most data were retrieved 
      from JPL's SBDB and HORIZONS using tools provided by the Python package Astroquery \citep{2019AJ....157...98G}. Although the 
      orbit determination still needs to be improved (the orbital size and shape are relatively good, but the orientation in space 
      is still somewhat uncertain), particularly when compared with that of 29P in Table~\ref{elements}, we believe that it is 
      good enough to reach some robust conclusions given the nature of our calculations. In addition to studying some 
      representative orbits, we performed longer calculations that applied the Monte Carlo using the Covariance Matrix (MCCM) 
      methodology described by \citet{2015MNRAS.453.1288D} in which a Monte Carlo process generates control or clone orbits (3000) 
      based on the nominal orbit, but adding random noise on each orbital element by making use of the covariance matrix, which 
      was also retrieved from JPL's SSDG SBDB. 

      \subsection{Orbital context}
         Here, we focus on the sample of objects with $a\in(5.4, 7)$~AU, $e<0.15$, and $i<10${\degr} that includes the four objects 
         in Table~\ref{elements} and eight other known small bodies that follow this type of low-eccentricity, low-inclination 
         orbit located just beyond that of Jupiter: 2011~FS$_{53}$, 2012~BS$_{76}$, 2014~EB$_{132}$, 2014~EF$_{115}$, 
         2014~EM$_{120}$, 2014~EO$_{68}$, 2014~EW$_{77}$, and 2016~AK$_{51}$. Unfortunately, all of them have very poor orbit 
         determinations based on about a dozen observations and spanning data arcs of 2 to 18 days. The distribution of mutual 
         nodal distances (their absolute values) was computed as described in Appendix~A, using data from JPL's SSDG SBDB. A 
         small mutual nodal distance implies that the objects might experience close flybys, but this must be confirmed by using 
         $N$-body simulations.

\begin{figure}
       \centering
        \includegraphics[width=\linewidth]{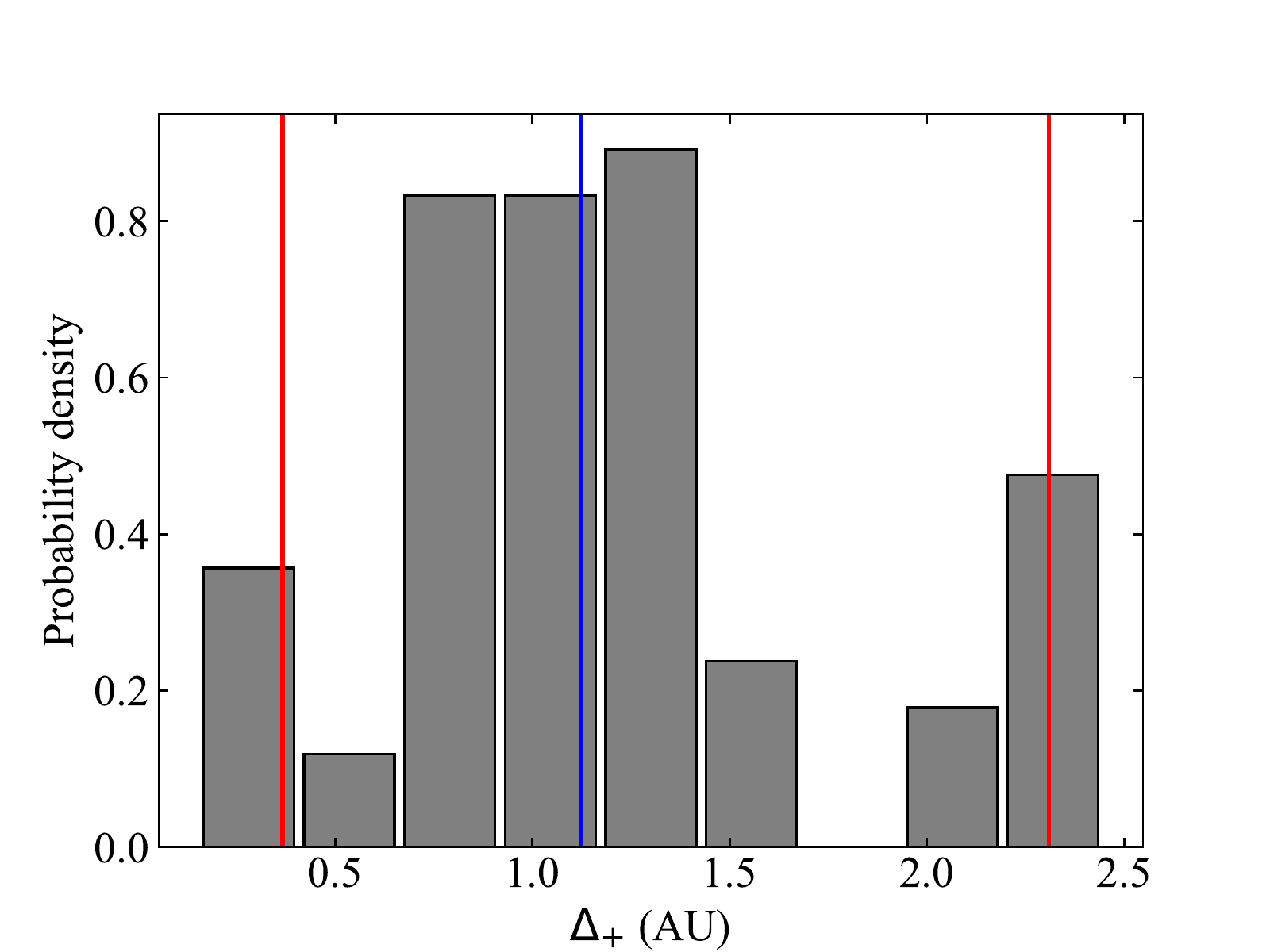}
        \includegraphics[width=\linewidth]{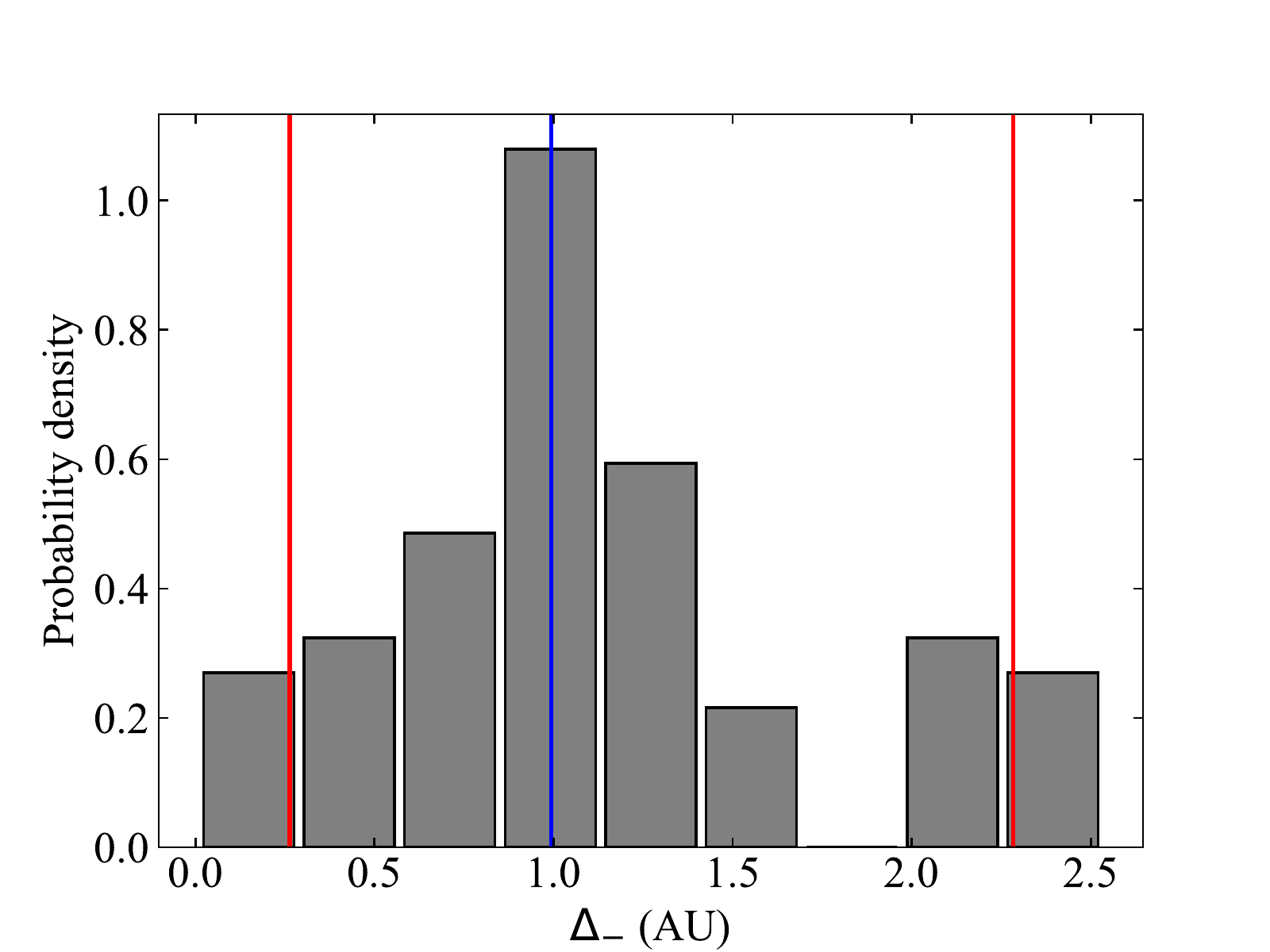}
        \caption{Distribution of mutual nodal distances. {\em Upper panel}: For the ascending mutual nodes of the sample of 12 
                 small bodies following orbits similar to that of 29P. The median is shown in blue and the 5th and 95th 
                 percentiles are in red. {\em Bottom panel}: For the descending mutual nodes of the same sample. In the histogram, 
                 we use bins computed using the Freedman and Diaconis rule \citep{Freedman1981} and counts to form a probability 
                 density so the area under the histogram will sum to one.
                }
        \label{ascdescnodes}
     \end{figure}

         Our sample produced 66 pairs of mutual nodal distances, ${\Delta}_{\pm}$ (the results for each pair come from a set of 
         10$^4$ pairs of virtual objects as described in Appendix~\ref{29Plikeelements}). The distribution of mutual nodal 
         distances for the ascending mutual nodes is shown in the upper panel of Fig.~\ref{ascdescnodes} and the one corresponding 
         to the descending mutual nodes is displayed in the bottom panel of Fig.~\ref{ascdescnodes}. These distributions were 
         computed using mean values and uncertainties in the orbit determinations as described in Appendix~\ref{29Plikeelements}. 
         The first percentile of the distribution in ${\Delta}_{+}$ is equal to 0.22~AU and the one of ${\Delta}_{-}$ is 0.04~AU. 
         The first percentile is often considered as the statistically significant boundary to select severe outliers. 
%
%
%
%

\begin{figure}
       \centering
        \includegraphics[width=\linewidth]{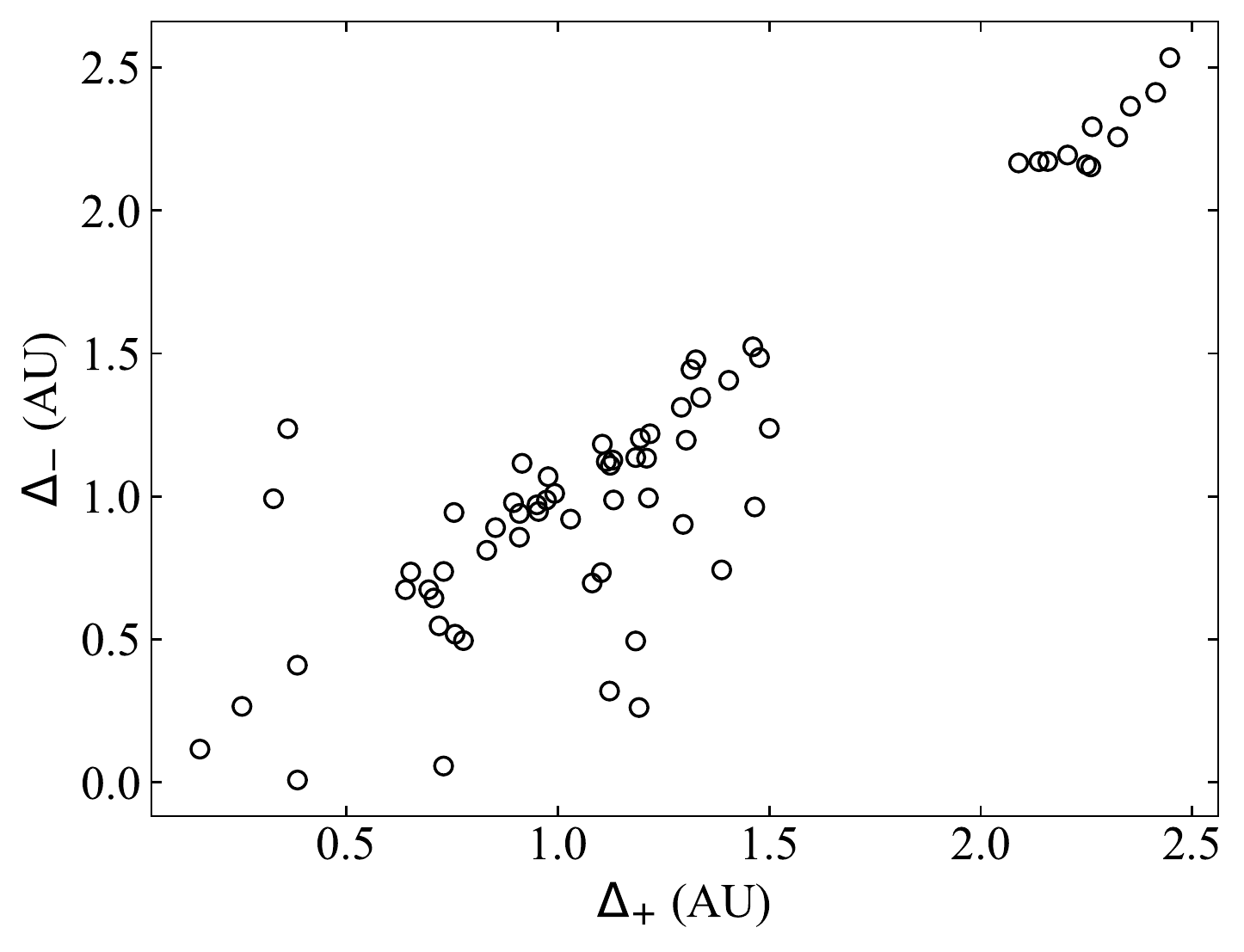}
        \includegraphics[width=\linewidth]{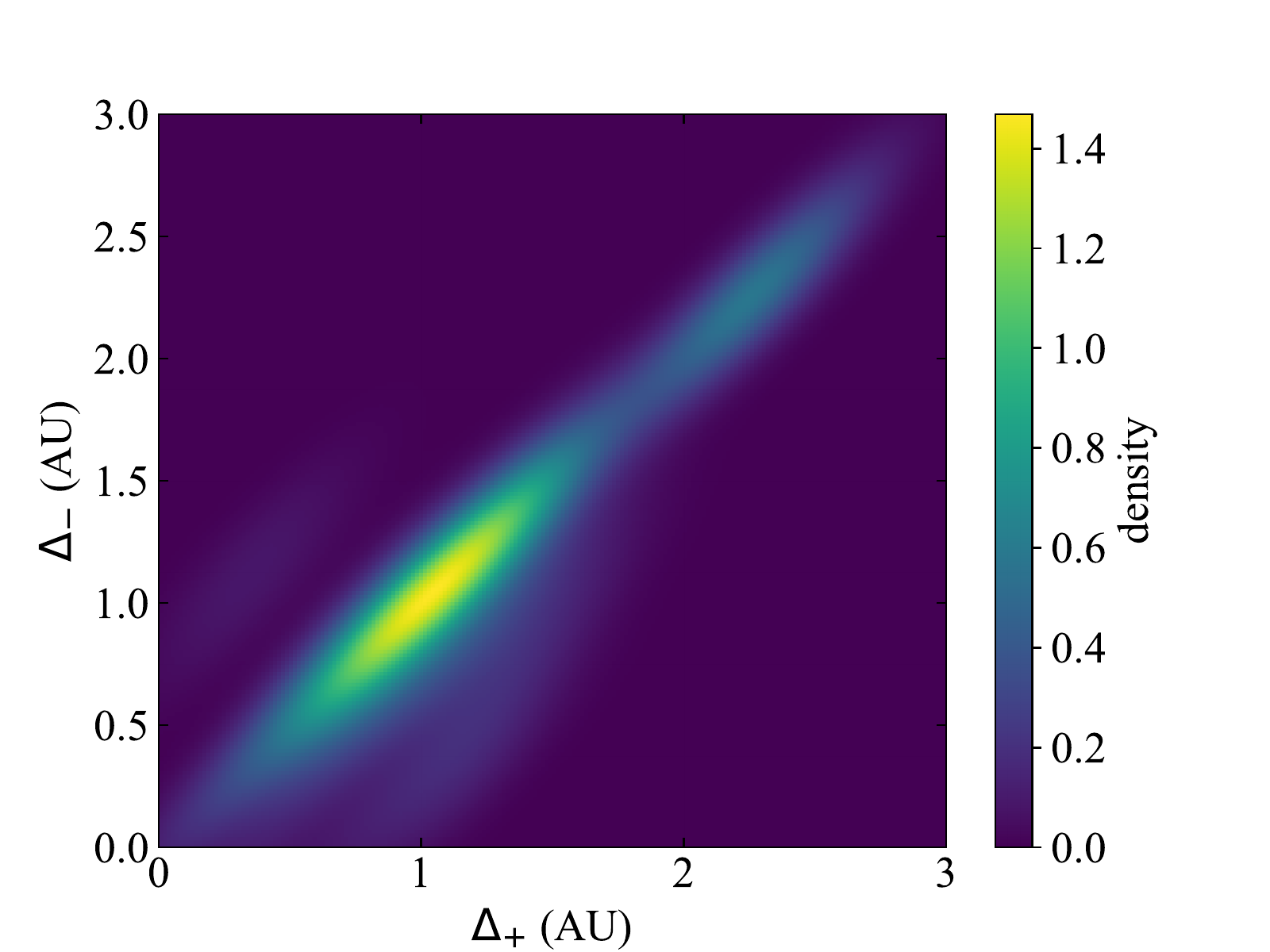}
        \caption{Mutual nodal distances of objects in 29P-like orbits. {\em Upper panel}: 66 pairs of mutual nodal distances. 
                 {\em Bottom panel}: Gaussian kernel density estimation of the same data.
                }
        \label{nodemap}
     \end{figure}

\begin{figure}
       \centering
        \includegraphics[width=\linewidth]{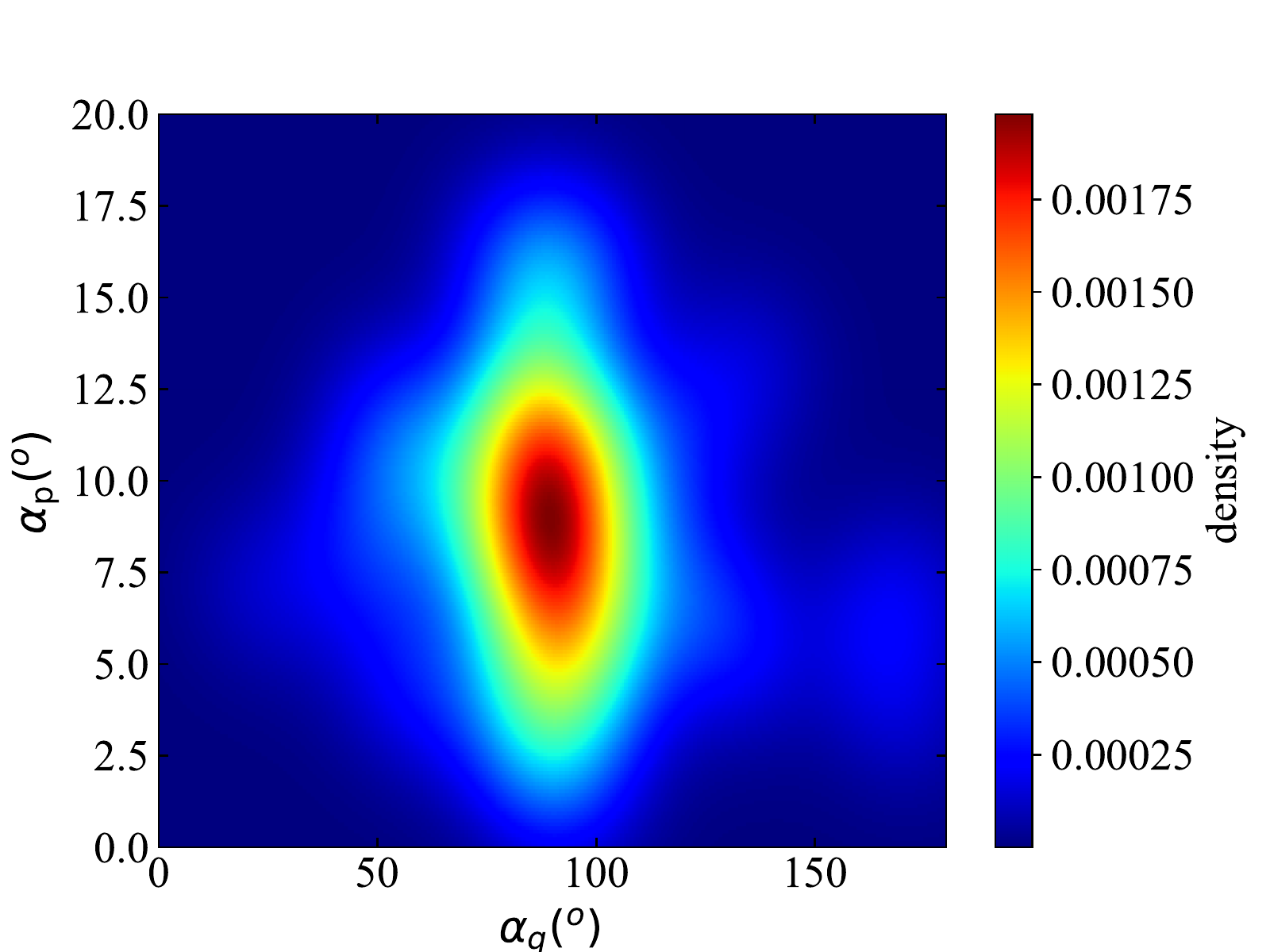}
        \includegraphics[width=\linewidth]{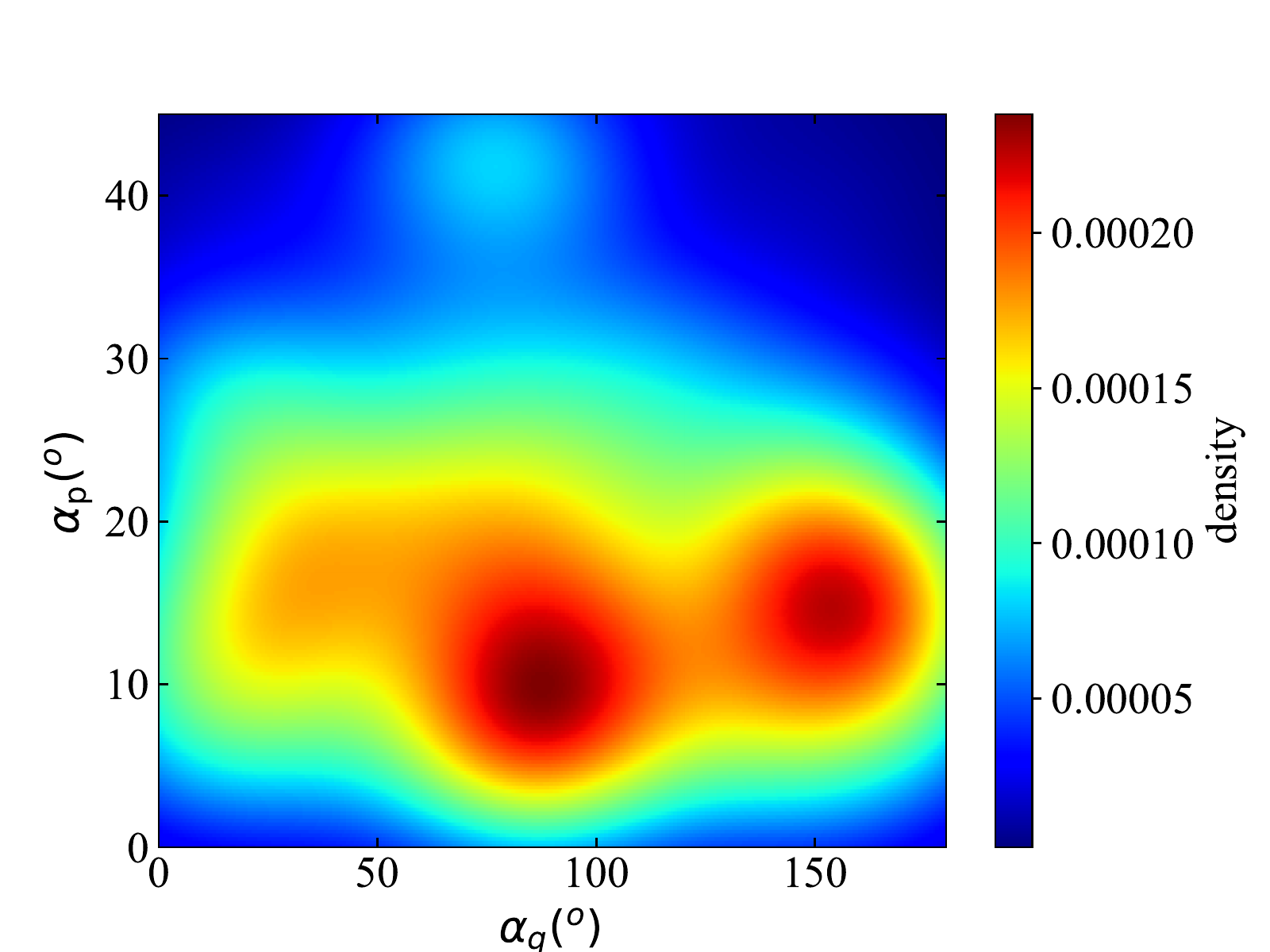}
        \caption{Orientations in space of the orbits. {\em Upper panel}: Gaussian kernel density estimation for the 66 pairs of
                 29P-like orbits. {\em Bottom panel}: Gaussian kernel density estimation for the 171 pairs of NC orbits.
                }
        \label{orbinspace}
     \end{figure}

         When considering the 66 pairs of mutual nodal distances, two clear outliers emerge: For 29P and 2020~MK$_{4}$ 
         ${\Delta}_{+}=0.1536\pm0.0005$~AU (median, and the 16th and 84th percentiles) and for 29P and P/2010~TO20 (LINEAR-Grauer) 
         ${\Delta}_{-}=0.0083\pm0.0003$~AU. In both cases, it is statistically unlikely that the small values of the mutual nodal 
         distances could be accidental (see Fig.~\ref{nodemap}) and some type of connection must exist, be it in the form of 
         resonant forces or a true physical relationship in which both objects come from a disrupted parent body. If disruption 
         events are the cause of the small mutual nodal distances, two of them may be required to explain the observed values. The 
         distributions of angular distances between pairs of orbital poles and perihelia were computed as described in 
         Appendix~B using data from JPL's SSDG SBDB. The orientations in space of the orbits of these objects are compatible with 
         those coming from a continuous uniform distribution (average and standard deviation values, see Fig.~\ref{orbinspace}, 
         upper panel) of angular distances between pairs of orbital poles and perihelia.
%
%
%
%
%
%
%
%

         It may be argued that our choice of parameter boundaries to select the sample of minor bodies that follow 29P-like orbits
         is somewhat artificial. \citet{2021Icar..35814201R} have studied the dynamics of small bodies that go around the Sun 
         between the orbits of Jupiter and Saturn. In their work, it is argued that all of these bodies have orbits similar to that 
         of comet 29P. They call this group the ``near centaurs" or NCs that have values of the perihelion distance $q>5.204$~AU 
         and the aphelion distance $Q\in(5.6, 9.583)$~AU. JPL's SSDG SBDB shows that this group includes 42 objects; after discarding 
         all the fragments of comet D/1993~F2 (Shoemaker-Levy 9), and 2004~VP$_{112}$ and 2007~TB$_{434}$ because their orbit 
         determinations are very uncertain, our working sample includes 19 objects. This NC sample does not include P/2010~TO20 
         (LINEAR-Grauer) nor most of the objects in our previous sample with the exception of 2011~FS$_{53}$, because their 
         perihelia are shorter, but it does include the other three objects in Table~\ref{elements}.  

\begin{figure}
       \centering
        \includegraphics[width=\linewidth]{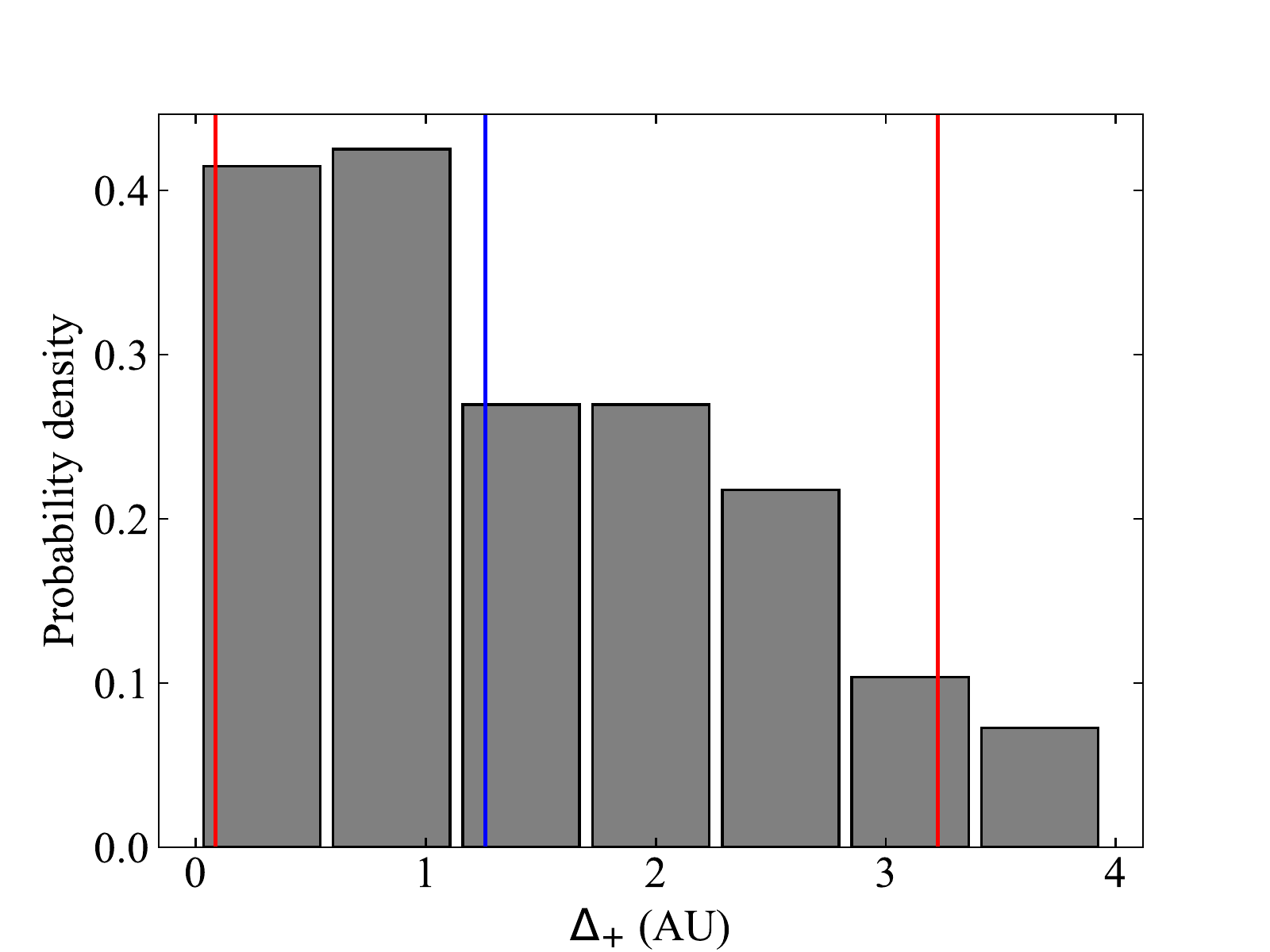}
        \includegraphics[width=\linewidth]{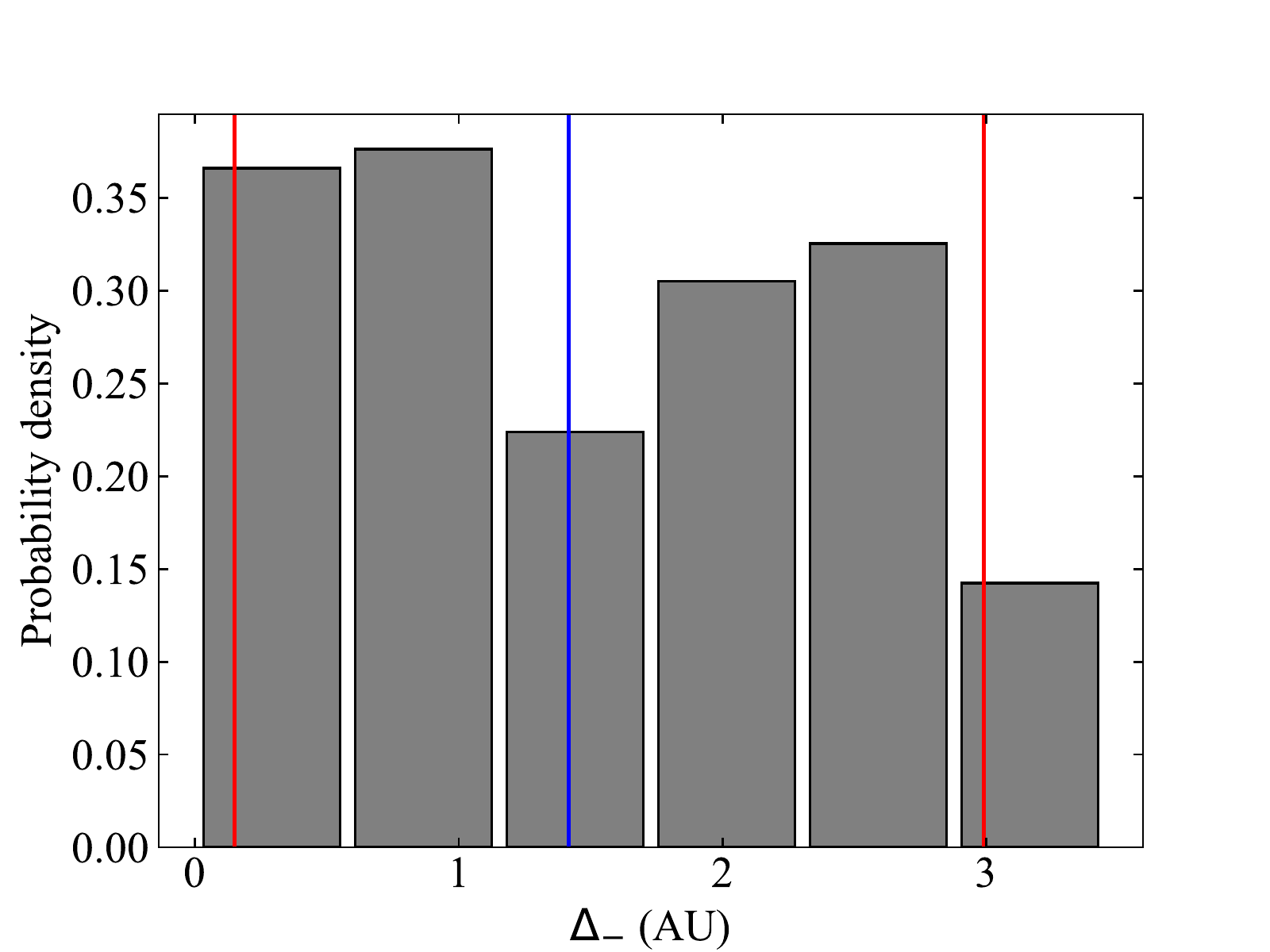}
        \caption{Distribution of mutual nodal distances. {\em Upper panel}: For the ascending mutual nodes of the sample of 19 
                 NCs. The median is shown in blue and the 5th and 95th percentiles are in red. {\em Bottom panel}: For the 
                 descending mutual nodes of the same sample. In the histogram, we use bins computed using the Freedman and 
                 Diaconis rule \citep{Freedman1981} and counts to form a probability density so the area under the histogram will 
                 sum to one.
                }
        \label{ascdescnodesNCs}
     \end{figure}

\begin{figure}
       \centering
        \includegraphics[width=\linewidth]{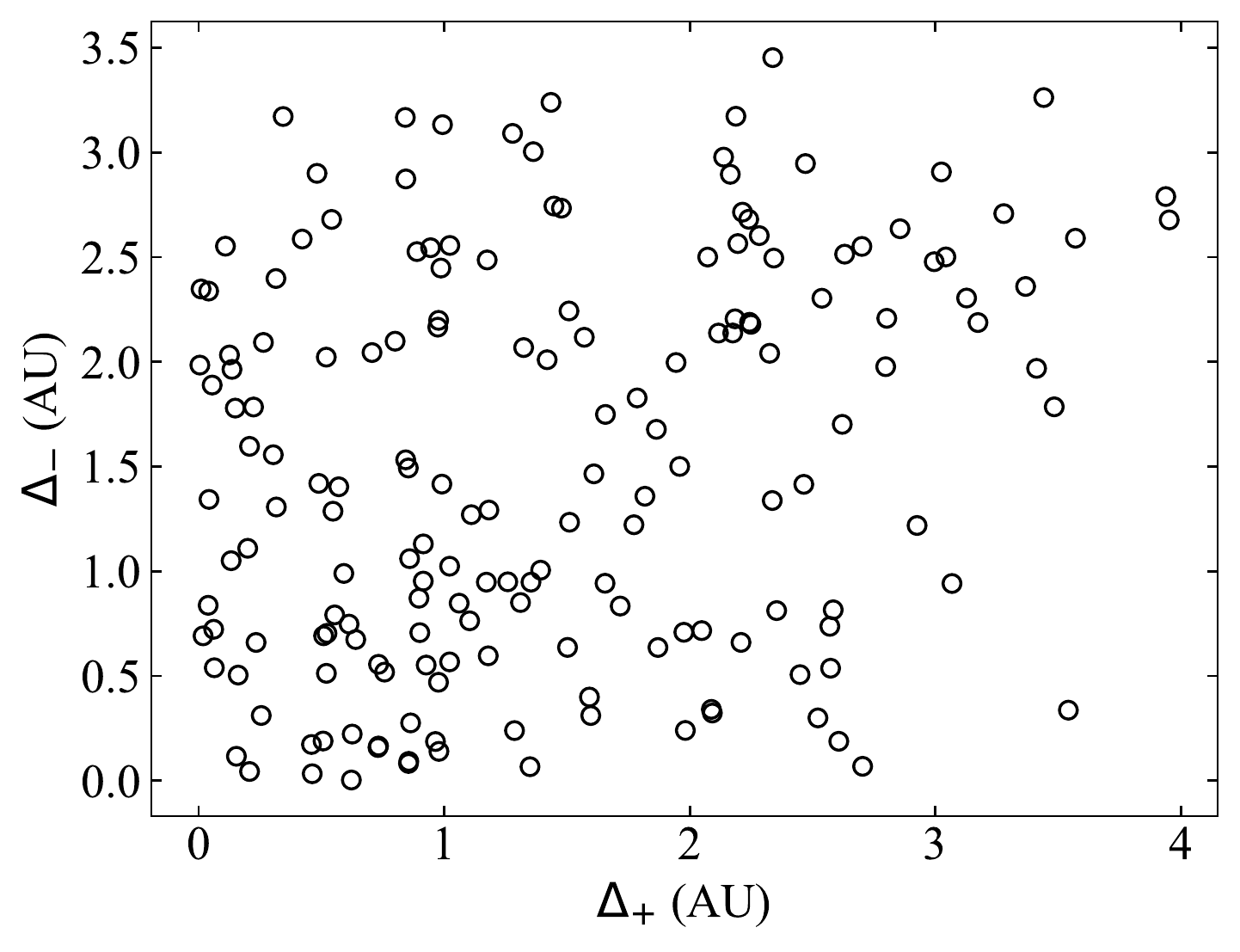}
        \includegraphics[width=\linewidth]{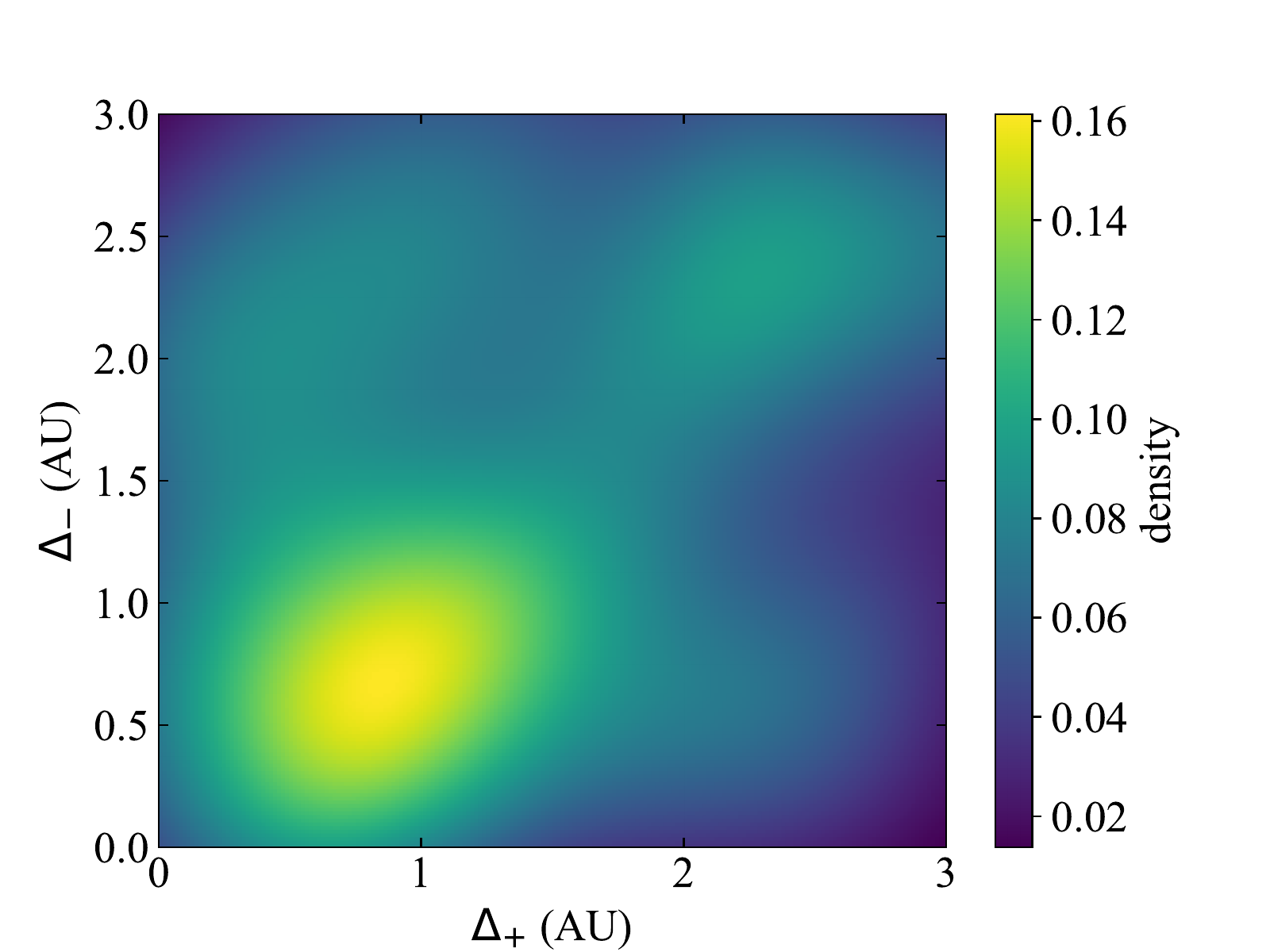}
        \caption{Mutual nodal distances of NCs. {\em Upper panel}: 171 pairs of mutual nodal distances. {\em Bottom panel}: 
                 Gaussian kernel density estimation of the same data.
                }
        \label{nodemapNCs}
     \end{figure}

         If we repeat the previous analysis for these NCs, we obtain Figs.~\ref{ascdescnodesNCs} and \ref{nodemapNCs}. For this 
         sample, the first percentile of the distribution in ${\Delta}_{+}$ is equal to 0.015~AU and the one of ${\Delta}_{-}$ is 
         0.040~AU. When considering the 171 pairs of mutual nodal distances (see Fig.~\ref{nodemapNCs}), four clear outliers 
         emerge: For 29P and (494219) 2016~LN$_{8}$ ${\Delta}_{-}=0.003090\pm0.000002$~AU, for 2015~UH$_{67}$ and P/2005~T3 (Read) 
         ${\Delta}_{-}=0.03\pm0.02$~AU, for P/2008~CL94 (Lemmon) and P/2011~C2 (Gibbs) ${\Delta}_{+}=0.0096\pm0.0005$~AU, and for 
         P/2008~CL94 (Lemmon) and P/2015~M2 (PANSTARRS) ${\Delta}_{+}=0.0047\pm0.0004$~AU. Although \citet{2021Icar..35814201R} 
         argue that most of these objects are partly subjected to various mean-motion resonances with the giant planets, such 
         small values of ${\Delta}_{\pm}$ are suggestive of an interacting population in which fragmentation events may be taking 
         place during outburst episodes. 
%
%
%
%
%
%
%
%

         Additional evidence along this line of reasoning comes from the study of the distributions of angular distances between 
         pairs of orbital poles and perihelia defined by the angles $\alpha_{\rm p}$ and $\alpha_{q}$ (computed as described in
         Appendix~B, see Fig.~\ref{orbinspace}, bottom panel), and the difference in time of perihelion passage, $\Delta{T_{q}}$. 
         The first percentiles of these distributions are 2\fdg68, 5\fdg64, and 0.062~yr, respectively. Outlier pairs are as 
         follows: 2013~SO$_{107}$ and 2014~HY$_{195}$ with $\alpha_{\rm p}=2\fdg252\pm0\fdg003$, 39P/Oterma and P/2005~S2 (Skiff) 
         with $\alpha_{q}=5\fdg51\pm0\fdg12$, 39P/Oterma and P/2005~T3 (Read) $\alpha_{q}=1\fdg2_{-0\fdg3}^{+0\fdg5}$, P/2005~T3 
         (Read) and P/2008~CL94 (Lemmon) with $\alpha_{\rm p}=2\fdg236\pm0\fdg011$, 2015~UH$_{67}$ and 2020~MK$_{4}$ with 
         $\Delta{T_{q}}=0.013_{-0.008}^{+0.010}$~yr, and 2020~MK$_{4}$ and P/2015~M2 (PANSTARRS) with 
         $\Delta{T_{q}}=0.053\pm0.010$~yr. Some of these objects have highly correlated orbits in terms of their orientation in 
         space and timing, which are difficult to explain by chance or mean-motion resonances alone. 

      \subsection{Current dynamical status}
         Minor body 2020~MK$_{4}$ goes around the Sun between the orbits of Jupiter and Neptune, so it is a centaur. It has a 
         current value of the Tisserand's parameter, $T_{\rm J}$ \citep{1999ssd..book.....M}, of 3.005; therefore, and following 
         \citet{1997Icar..127...13L}, it cannot be a Jupiter-family comet because the value is not in the interval (2,~3), even if 
         we account for the uncertainties. In contrast, 29P has a value of the Tisserand parameter of 2.984 and it has remained 
         consistently under 3.0 since its discovery back in 1927. In this context, the Tisserand parameter, which is a 
         quasi-invariant, is given by the expression:
         \begin{equation}
            T_{\rm J} = \frac{a}{a_{\rm J}} + 2 \ \cos{i} \ \sqrt{\frac{a_{\rm J}}{a} \ (1 - e^{2})} \,, \label{Tisserand}
         \end{equation}
         where $a$, $e$, and $i$ are the semimajor axis, eccentricity, and inclination of the orbit of the minor body under study, 
         respectively, and $a_{\rm J}$ is the semimajor axis of the orbit of Jupiter \citep{1999ssd..book.....M}. The functional 
         form of this parameter makes it robust against relatively large variations in the values of the relevant orbital 
         parameters, which helps its application to objects with rather chaotic orbital evolutions.

\begin{figure*}
        \centering
        \includegraphics[width=0.49\linewidth]{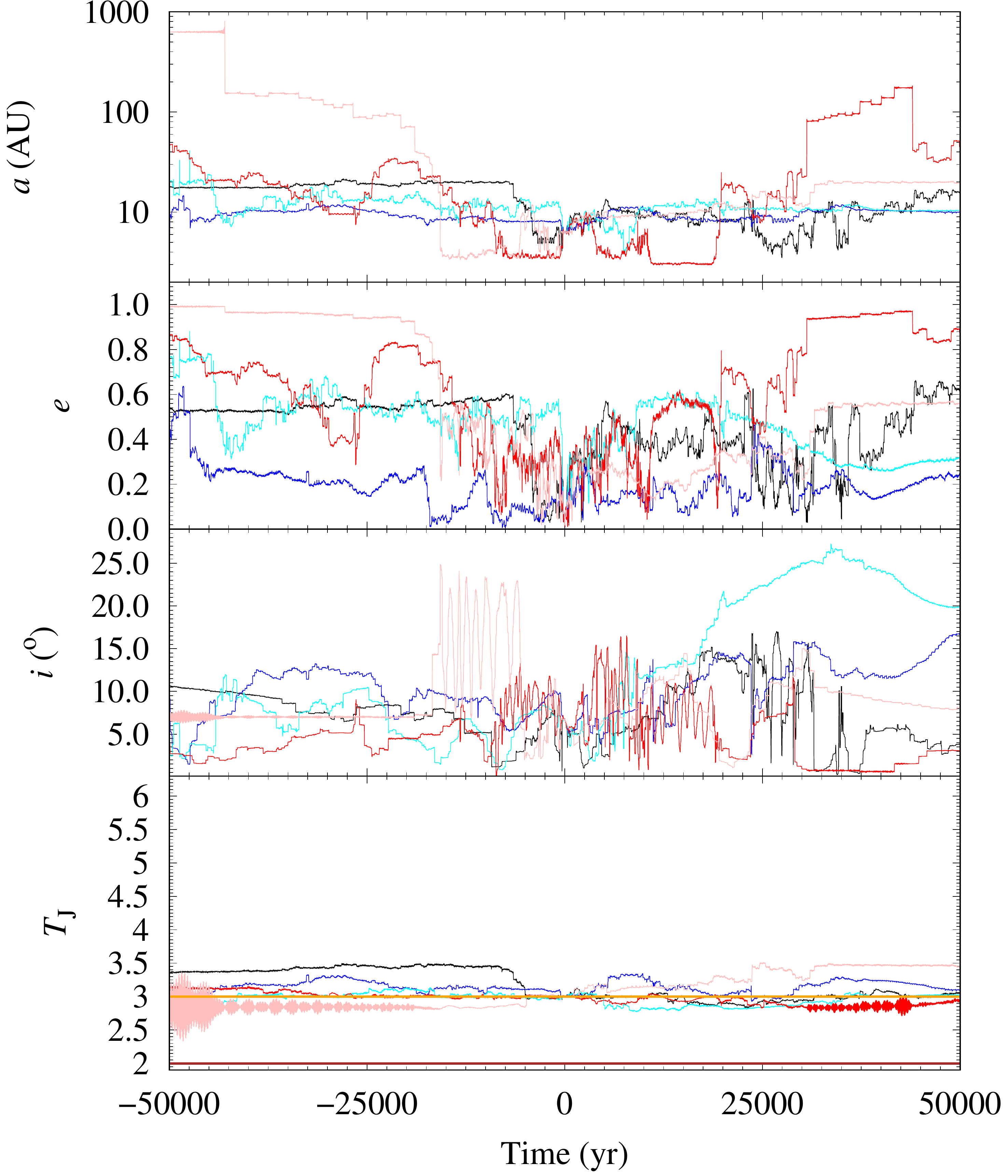}
        \includegraphics[width=0.49\linewidth]{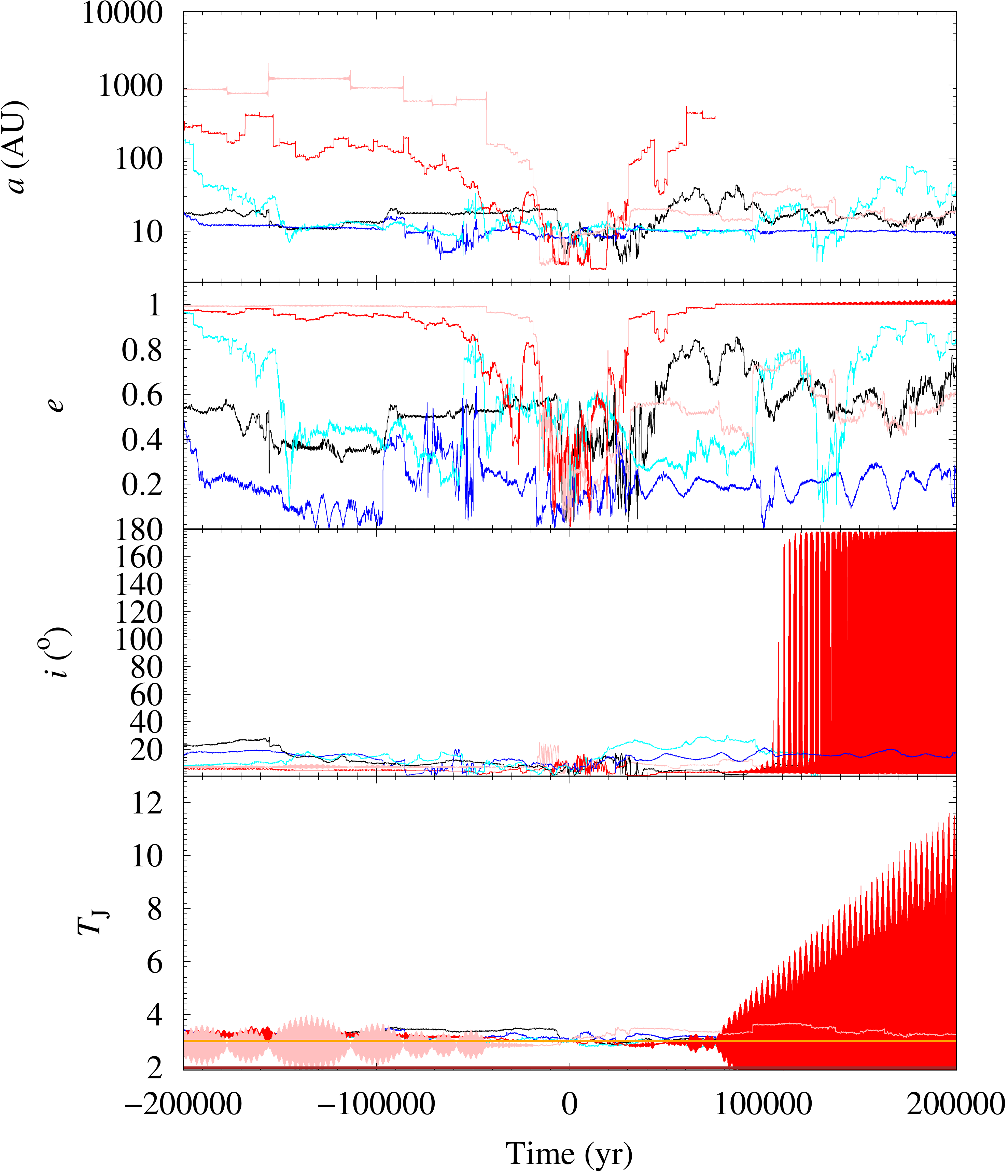}
        \caption{Short-term evolution of relevant parameters of 2020~MK$_{4}$. {\em Left panels}: focus on a shorter time window 
                 but correspond to the same data shown in the right-hand side panels. {\em Right panels}: evolution of the 
                 semimajor axis, $a$ (upper panels), of the nominal orbit (in black) as described by the orbit determination in 
                 Table~\ref{elements} and those of control orbits or clones with Cartesian vectors separated $+$3$\sigma$ (in 
                 blue), $-$3$\sigma$ (in cyan), $+$9$\sigma$ (in red), and $-$9$\sigma$ (in pink) from the nominal values in 
                 Table~\ref{vector}. The second from the top panels show the evolution of the eccentricity, $e$, for the same 
                 sample of control orbits. The second from the bottom panels display the inclination, $i$. The bottom panels show 
                 the variations of the Tisserand's parameter, $T_{\rm J}$ and include the boundary references 2 (in brown) and 3 
                 (in orange). The output time-step size is 20~yr, the origin of time is epoch 2459000.5 TDB.
                }
        \label{evolution}
     \end{figure*}

\begin{figure}
        \centering
        \includegraphics[width=\linewidth]{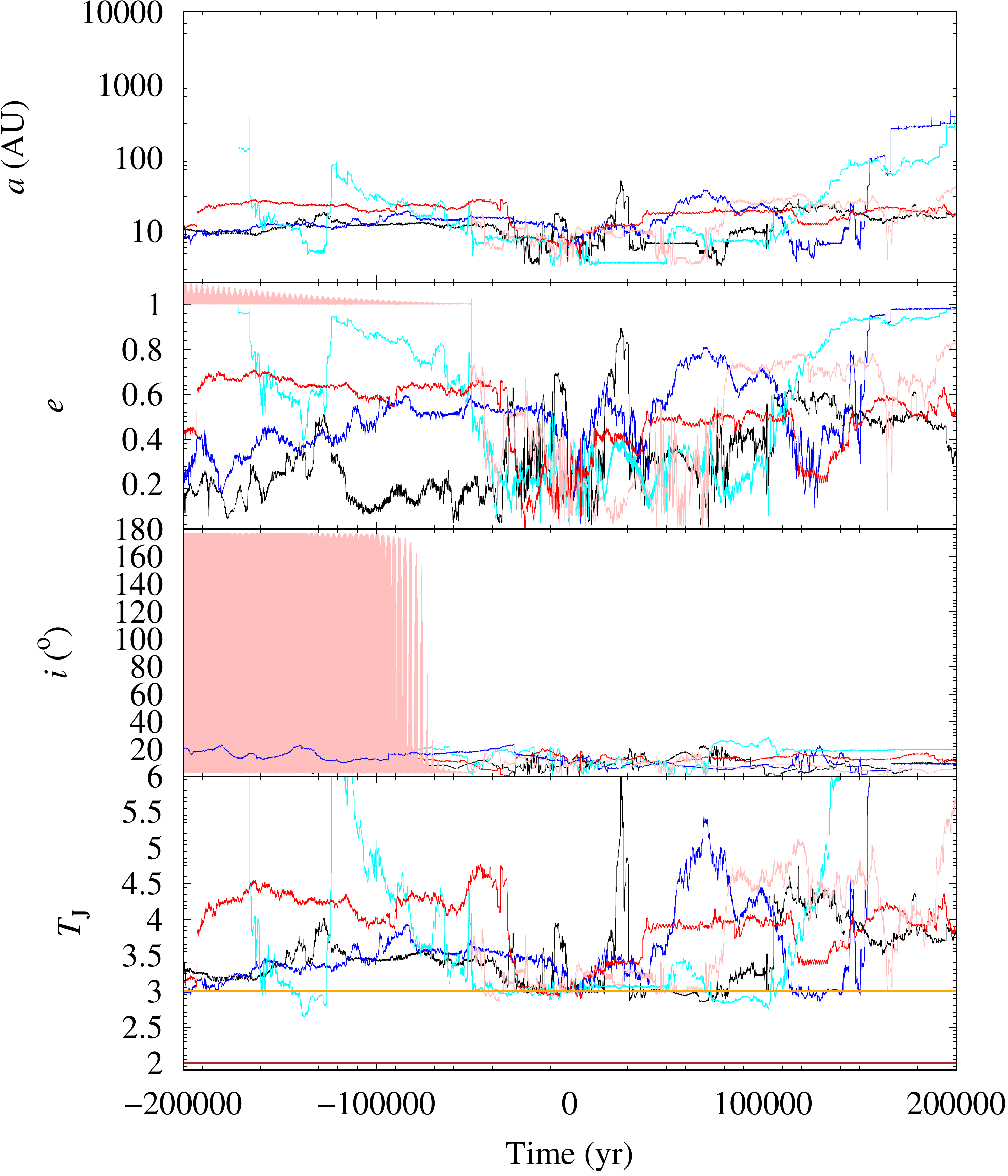}
        \caption{Short-term evolution of relevant parameters of comet 29P/Schwassmann-Wachmann~1. Similar to Fig.~\ref{evolution} 
                 but for data in Table~\ref{vector29P}. The output time-step size is 20~yr, the origin of time is epoch 2459000.5 
                 TDB.
                }
        \label{evolution29P}
     \end{figure}

         The panels on the right-hand side of Figure~\ref{evolution} show the short-term evolution of relevant parameters of 
         representative control orbits with Cartesian vectors separated by $\pm$3$\sigma$ and $\pm$9$\sigma$ from the nominal 
         values in Table~\ref{vector}. The orbital evolution is very chaotic and some instances lead to ejections integrating into 
         the past (not shown) and towards the future. Ejections are the result of very close encounters with Jupiter and other 
         giant planets, but also with the Sun following episodes similar to those described in \citet{2015MNRAS.446.1867D} for 
         comet 96P/Machholz~1. Close encounters with Jupiter drive the very chaotic short-term behavior observed in the panels on 
         the left in Fig.~\ref{evolution}. The evolution of $T_{\rm J}$ (bottom panels) shows that 2020~MK$_{4}$ is unlikely to 
         become a long-term member of the Jupiter-family comet dynamical class for control orbits in the $\pm$3$\sigma$ range, 
         both in the past and the future. However, control orbits more separated from the nominal one, $\pm$9$\sigma$, show 
         extended lengthy incursions inside the Jupiter-family comet orbital domain (the value of $T_{\rm J}$ librates, see bottom 
         panels in Fig.~\ref{evolution}). The bottom panel of Figure~\ref{evolution29P} shows that 29P has a much lower 
         probability of being a long-term member of the Jupiter-family comet group (the control orbits are based on the data in 
         Table~\ref{vector29P}). In general terms, the orbital evolution displayed in Figs.~\ref{evolution} and \ref{evolution29P} 
         is similar and nearly equally chaotic (readers are encouraged to compare the evolution of $a$, $e$, and $i$ in both 
         figures); this is the standard dynamical behavior for objects in these orbits (see for example 
         \citealt{2019MNRAS.490.4388G}).
%
%
%
%
%
%
%
%

\begin{figure*}
        \centering
        \includegraphics[width=0.49\linewidth]{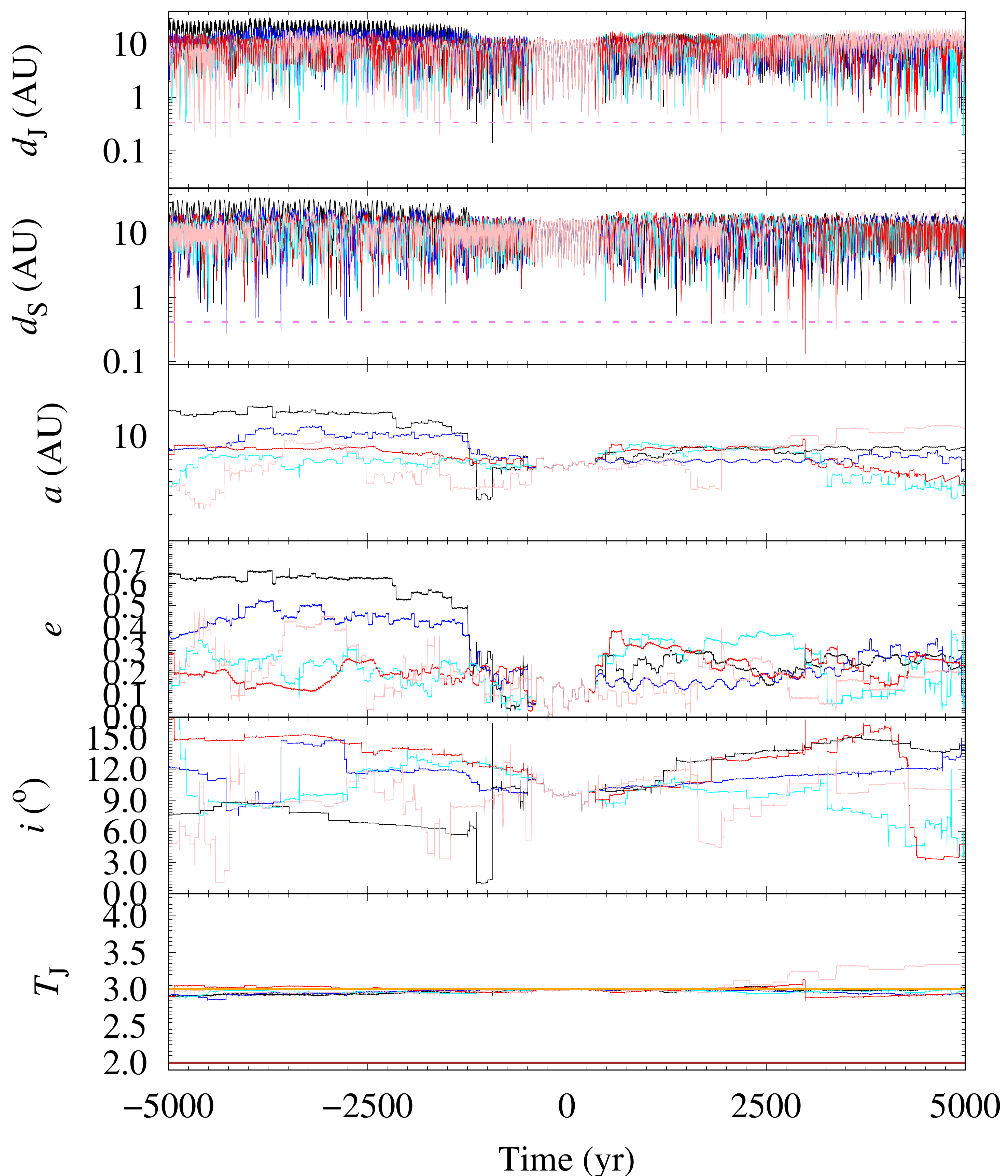}
        \includegraphics[width=0.49\linewidth]{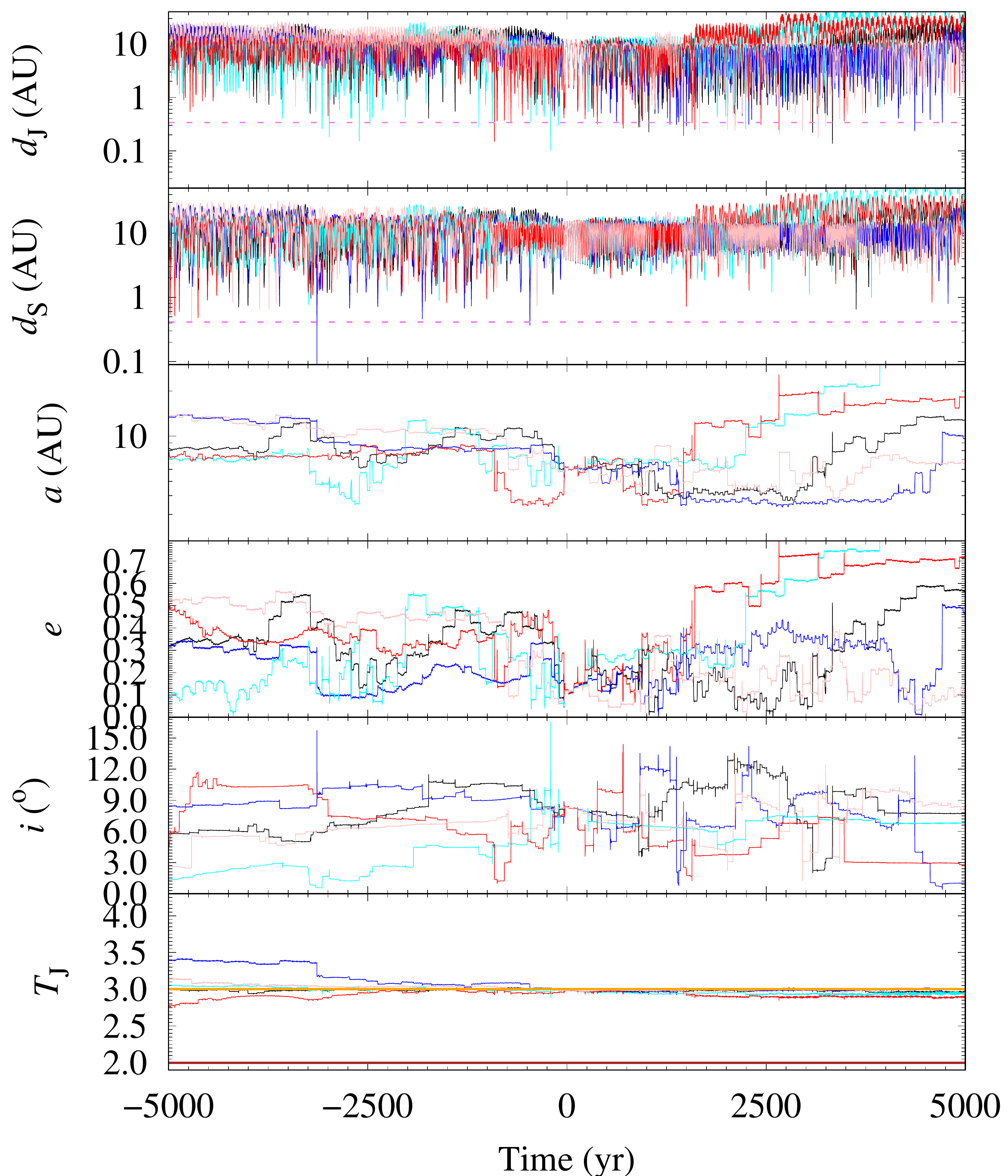}
        \includegraphics[width=0.49\linewidth]{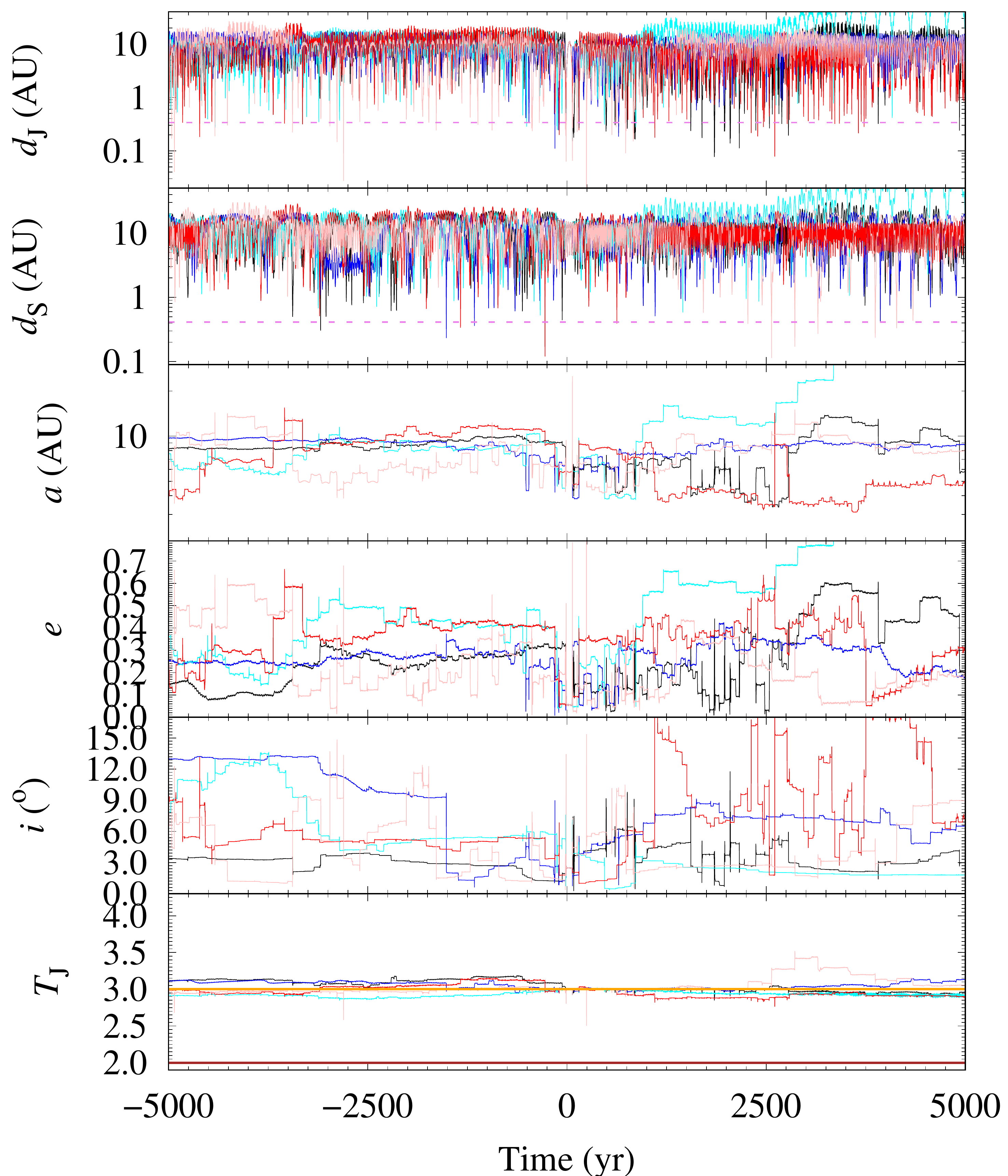}
        \includegraphics[width=0.49\linewidth]{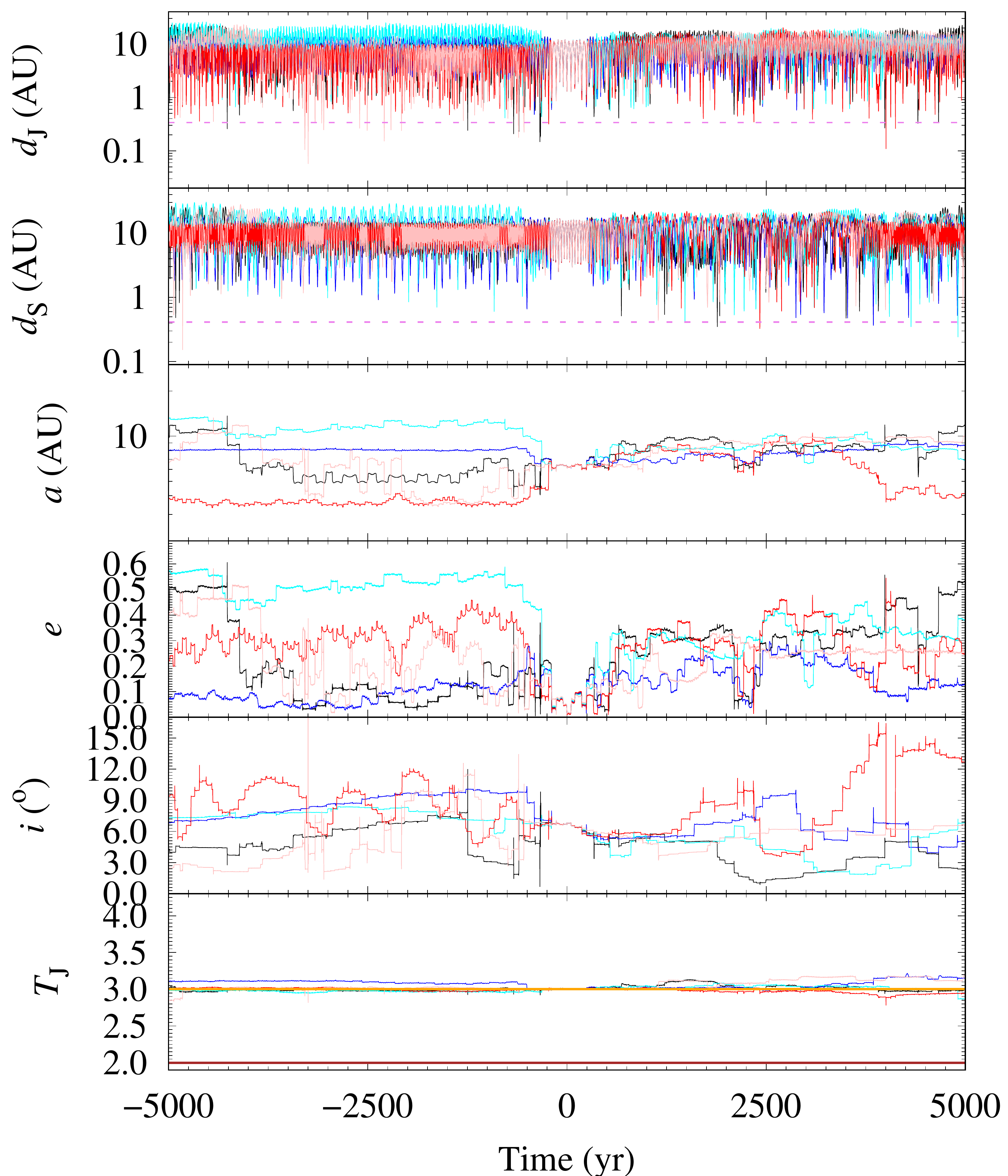}
        \caption{Short-term evolution of relevant parameters of comets 29P/Schwassmann-Wachmann~1, P/2008~CL94 (Lemmon), and 
                 P/2010~TO20 (LINEAR-Grauer), and centaur 2020~MK$_{4}$. Results for 29P are shown in the upper left panels, those 
                 of P/2008~CL94 are displayed in the upper right panels, the bottom left panels show those of P/2010~TO20, and the
                 bottom right panels focus on 2020~MK$_{4}$. For each set of panels, we show the evolution of the distances to 
                 Jupiter (top panel) and Saturn (second to top) of the nominal orbit (in black) as described by the corresponding 
                 orbit determination in Table~\ref{elements} and those of control orbits or clones with Cartesian vectors 
                 separated $+$3$\sigma$ (in blue), $-$3$\sigma$ (in cyan), $+$9$\sigma$ (in red), and $-$9$\sigma$ (in pink) from 
                 the nominal values in Appendix~\ref{Adata}. The Hill radii of Jupiter, 0.338~AU, and Saturn, 0.412~AU, is shown 
                 in red. The third panel from the top shows the evolution of the semimajor axis, $a$. The third panel from the 
                 bottom shows the evolution of the eccentricity, $e$, for the same sample of control orbits. The second panel 
                 from the bottom displays the inclination, $i$. The bottom panel shows the variations in Tisserand's parameter, 
                 $T_{\rm J}$, and includes the boundary references 2 (in brown) and 3 (in orange). The output time-step size is 
                 1~yr, and the origin of time is epoch 2459000.5 TDB.
                }
        \label{STevolution}
     \end{figure*}

         Figure~\ref{STevolution} shows the shorter-term evolution of relevant parameters of representative orbits for all the
         objects in Table~\ref{elements}. Both 29P and 2020~MK$_{4}$ are not currently experiencing very close encounters with
         Jupiter and Saturn, but P/2008~CL94 (Lemmon) approaches Jupiter inside the Hill radius of the planet and P/2010~TO20 
         (LINEAR-Grauer) does the same for both Jupiter and Saturn. 
%
%
%
%

      \subsection{Future orbital evolution}
\begin{figure}
        \centering
         \includegraphics[width=\linewidth]{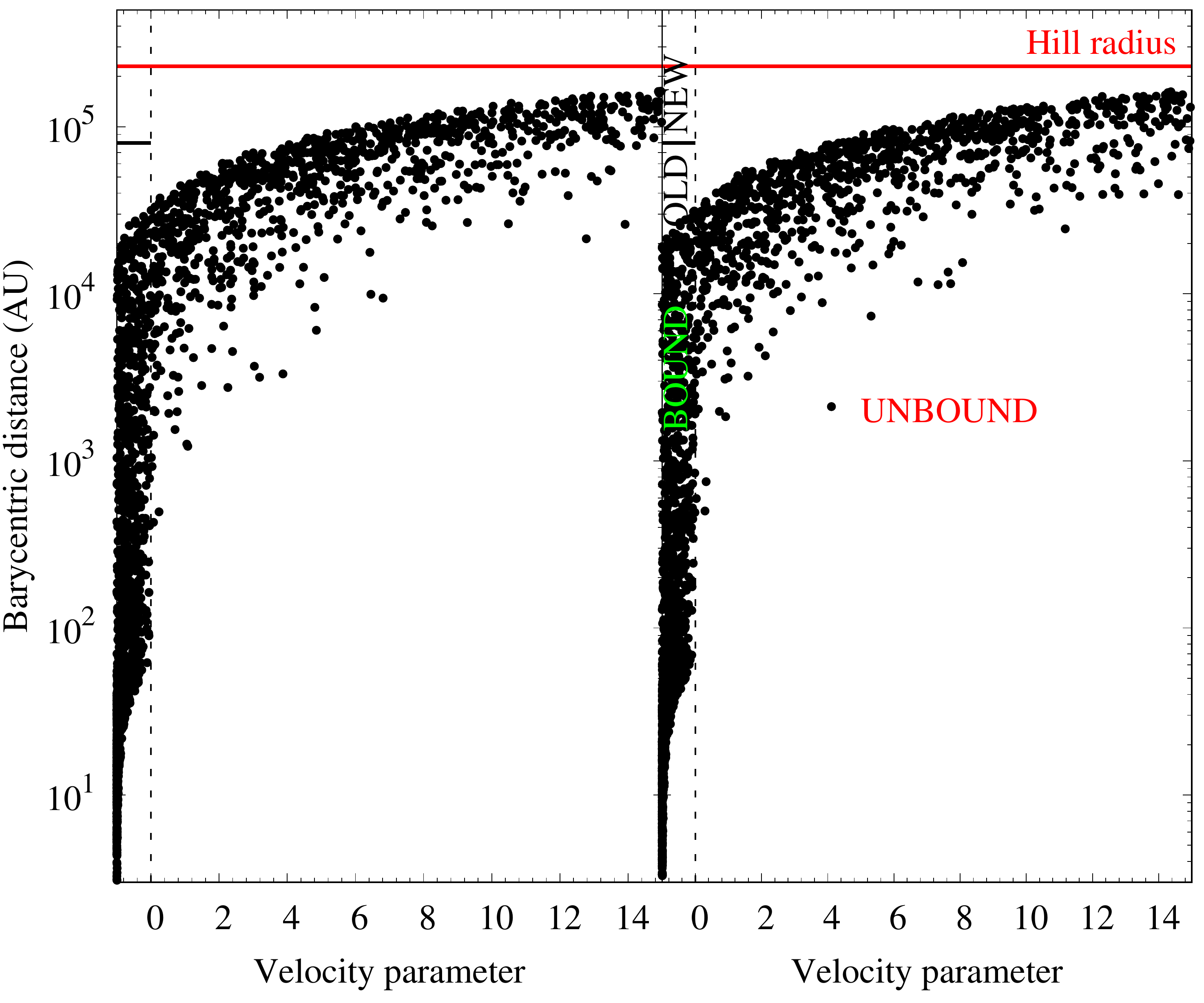}
         \caption{Past and future of 2020~MK$_{4}$. Values of the barycentric distance as a function of the velocity parameter 
                  0.5~Myr into the past (left panel) and future (right panel) for 3000 control orbits of 2020~MK$_{4}$. The 
                  velocity parameter is the difference between the barycentric and escape velocities at the computed barycentric 
                  distance in units of the escape velocity. Positive values of the velocity parameter identify control orbits that 
                  could be the result of capture (left panel) or lead to escape (right panel). The thick black line corresponds to 
                  the aphelion distance ---$a \ (1 + e)$, limiting case $e=1$--- that defines the domain of dynamically old comets 
                  with $a^{-1}>2.5\times10^{-5}$~AU$^{-1}$ (see \citealt{2017MNRAS.472.4634K}); the thick red line marks the 
                  radius of the Hill sphere of the Solar System (see for example \citealt{1965SvA.....8..787C}).
                 }
         \label{origin}
      \end{figure}

         Figure~\ref{evolution} shows that 2020~MK$_{4}$ is not unlikely to leave the Solar System within the next 200~kyr. 
         However, close encounters with Jupiter may lead to an eventual collision with the giant planet. A similar picture, but 
         less extreme, emerges for 29P when considering Fig.~\ref{evolution29P} and is consistent with the previous work
         by \citet{2017CoSka..47....7N}, \citet{2019ApJ...883L..25S}, and \citet{2021Icar..35814201R}. Longer calculations carried 
         out using MCCM to generate control orbits of 2020~MK$_{4}$ (see Fig.~\ref{origin}, right-hand side panel) yield a value 
         for the probability of ejection from the Solar System during the next 0.5~Myr of 0.48$\pm$0.03 (average and standard 
         deviation). These results are robust as they remain consistent between orbit determinations.
%
%
%
%

         The bottom right panels of Figure~\ref{STevolution} show that the future evolution of 2020~MK$_{4}$ becomes more chaotic 
         a few hundred years into the future when it will start experiencing close encounters under the Hill radius with Jupiter. 
         Nearly 700~yr into the future, it will start experiencing close encounters under the Hill radius with Saturn as well and 
         the overall orbital evolution will become even more chaotic. A similar behavior is observed for 29P (see 
         Fig.~\ref{STevolution}, upper left panels). P/2008~CL94 is far more engaged with Jupiter, interacting at a close range 
         now and in the future, but it will remain fairly detached from Saturn (see Fig.~\ref{STevolution}, upper right panels). 
         P/2010~TO20 remains strongly perturbed by Jupiter and is slightly less affected by Saturn (see Fig.~\ref{STevolution}, 
         bottom left panels), although close encounters with both planets seem to drive its very chaotic orbital evolution. All 
         these objects have a significant probability of leaving the Solar System in the relatively near future.        

      \subsection{Past orbital evolution: Possible origin}
         The key unknowns to determine are the origin of 2020 MK4 and whether 2020\ MK4 and 29P, and perhaps other objects, are 
         related. Figures~\ref{evolution} and \ref{evolution29P} show that there is a clear resemblance between the past orbital 
         evolution of both objects, but this does not imply a physical relationship as the evolutions are both very chaotic. An 
         exploration of this scenario requires the analysis of a large set of $N$-body simulations integrating backwards in time 
         for this pair. The statistical study of minimum approach distances may help in supporting or rejecting a scenario in 
         which 2020~MK$_{4}$ could be a fragment of 29P.

\begin{figure}
        \centering
         \includegraphics[width=\linewidth]{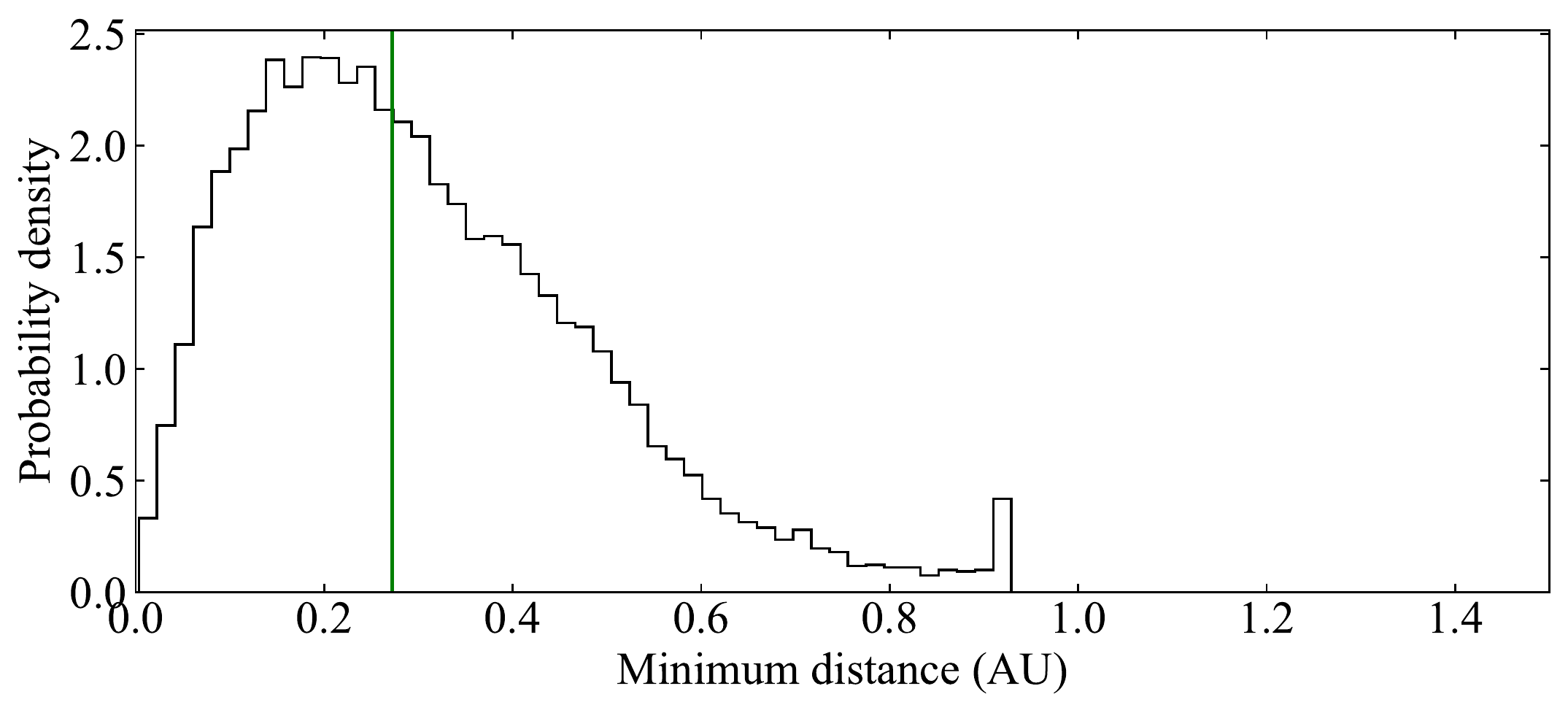}
         \includegraphics[width=\linewidth]{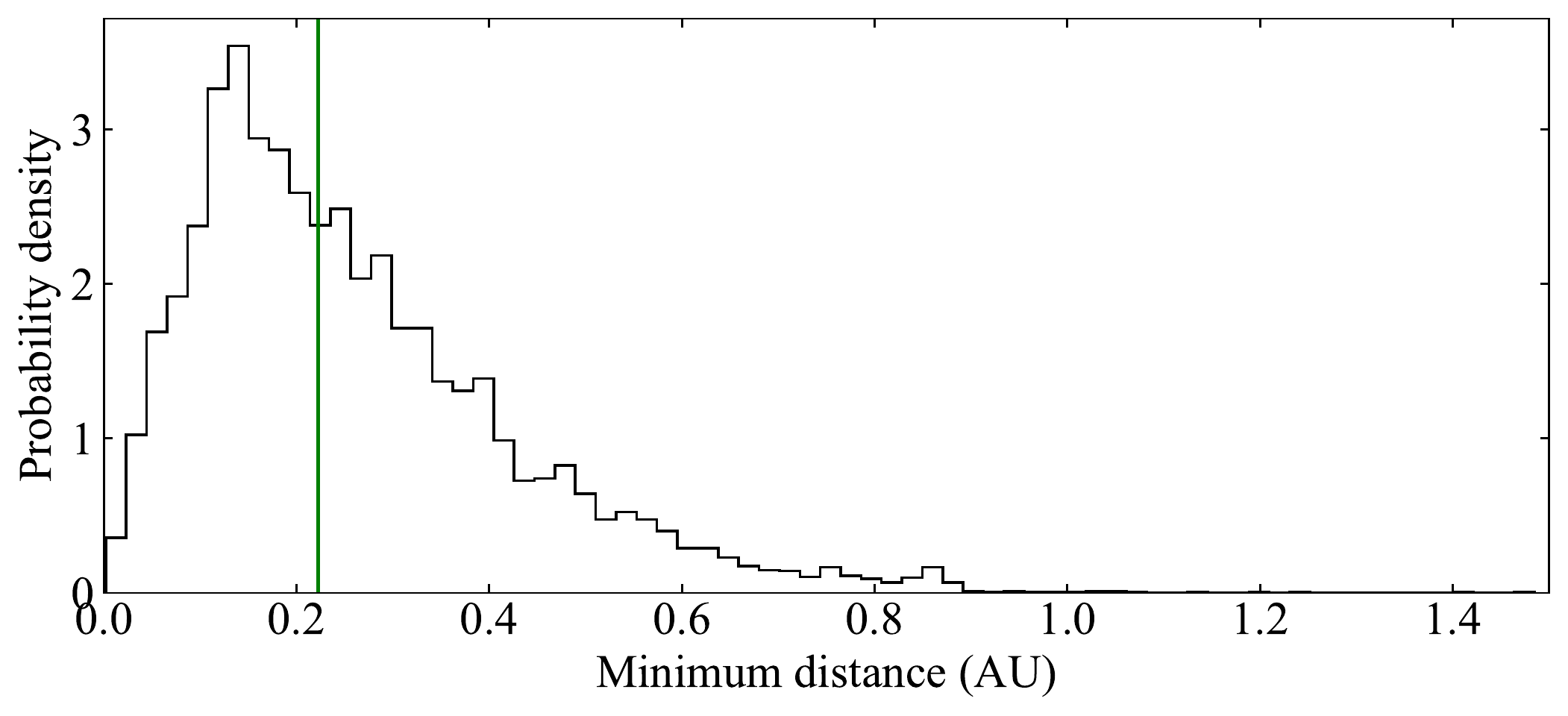}
         \includegraphics[width=\linewidth]{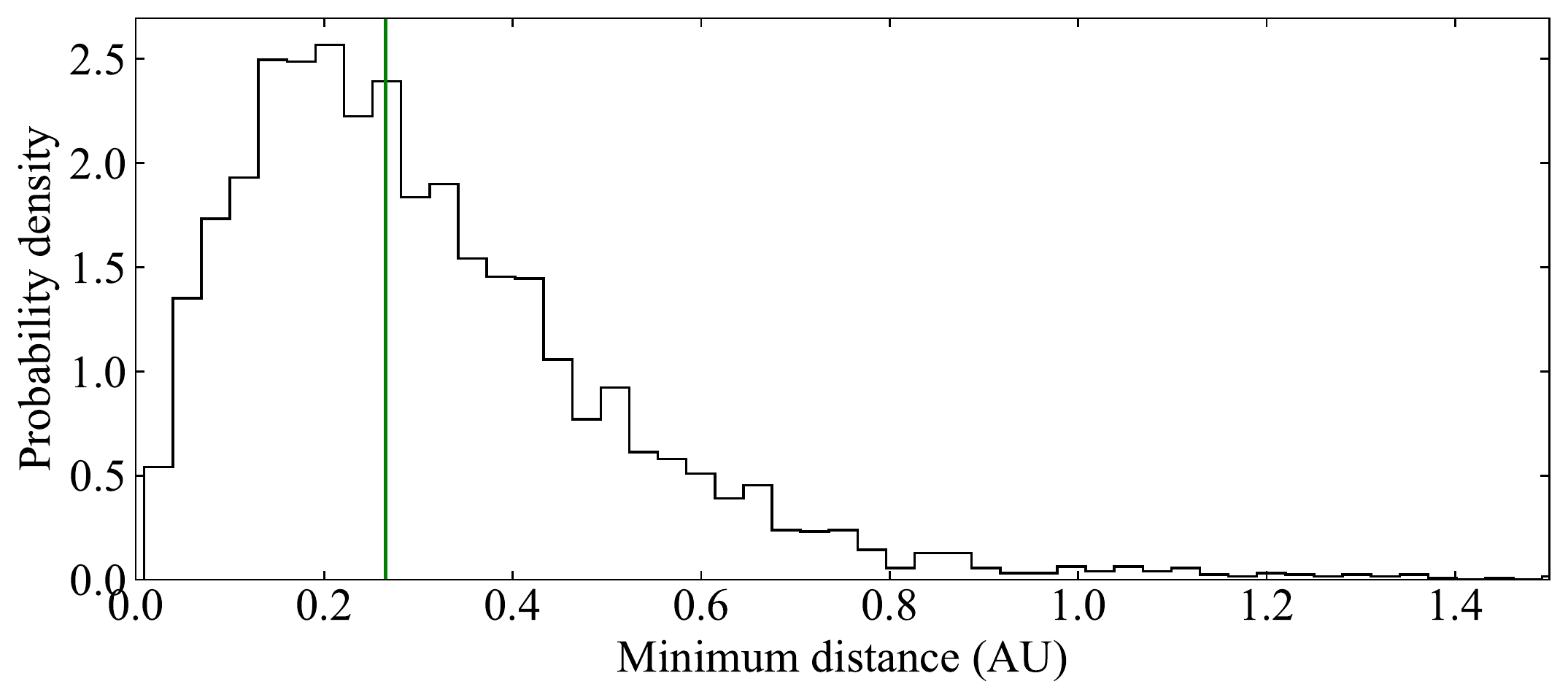}
         \caption{Distribution of minimum approach distances. {\em Upper panel:} For the pair 29P/Schwassmann-Wachmann~1 and 
                  2020~MK$_{4}$. {\em Middle panel:} For the pair P/2008~CL94 (Lemmon) and 2020~MK$_{4}$. {\em Bottom panel:} For 
                  the pair P/2010~TO20 (LINEAR-Grauer) and 2020~MK$_{4}$. Median values are displayed as vertical green lines. The 
                  bins were computed using the Freedman and Diaconis rule implemented in NumPy 
                  \citep{2011CSE....13b..22V,2020NumPy-Array}. In the histogram, we use counts to form a probability density so 
                  the area under the histogram will sum to one.
                 }
         \label{ddist}
      \end{figure}

\begin{figure}
        \centering
         \includegraphics[width=\linewidth]{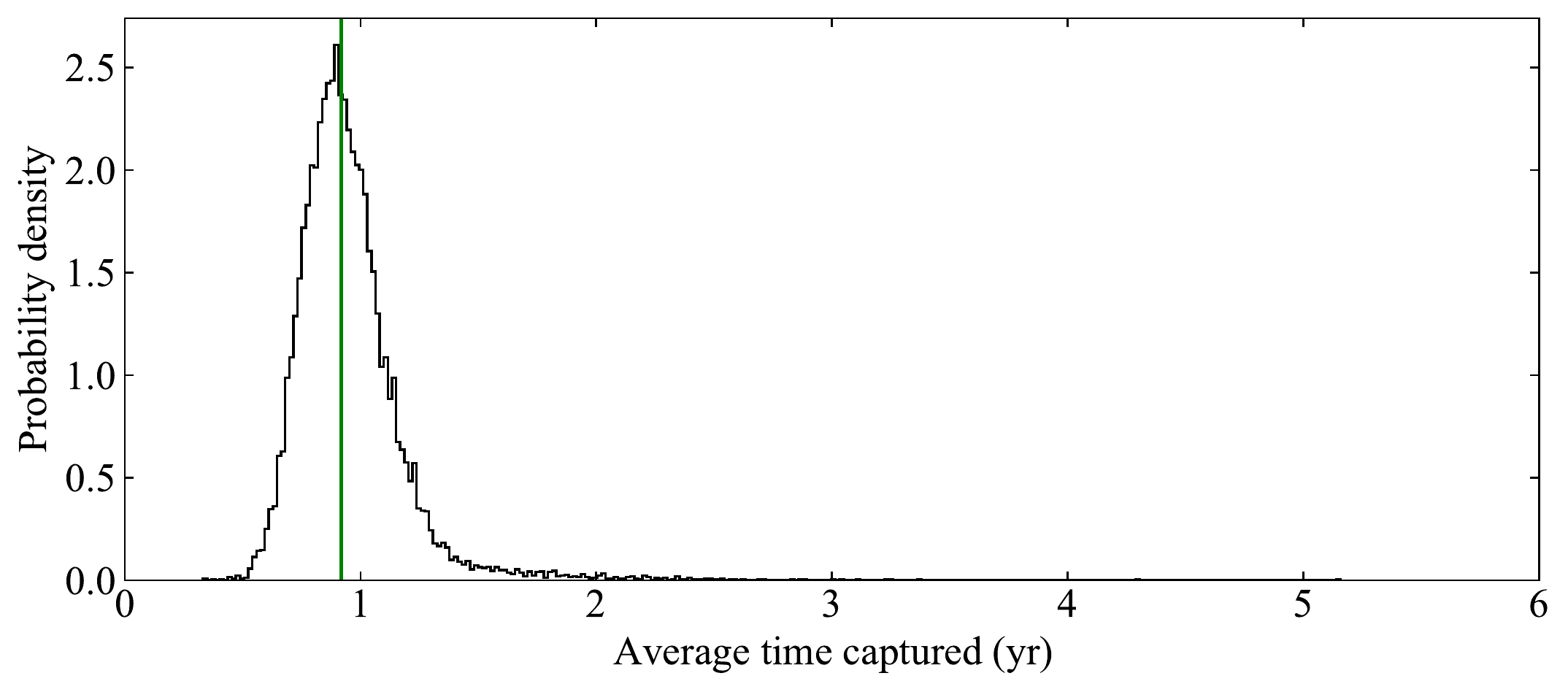}
         \includegraphics[width=\linewidth]{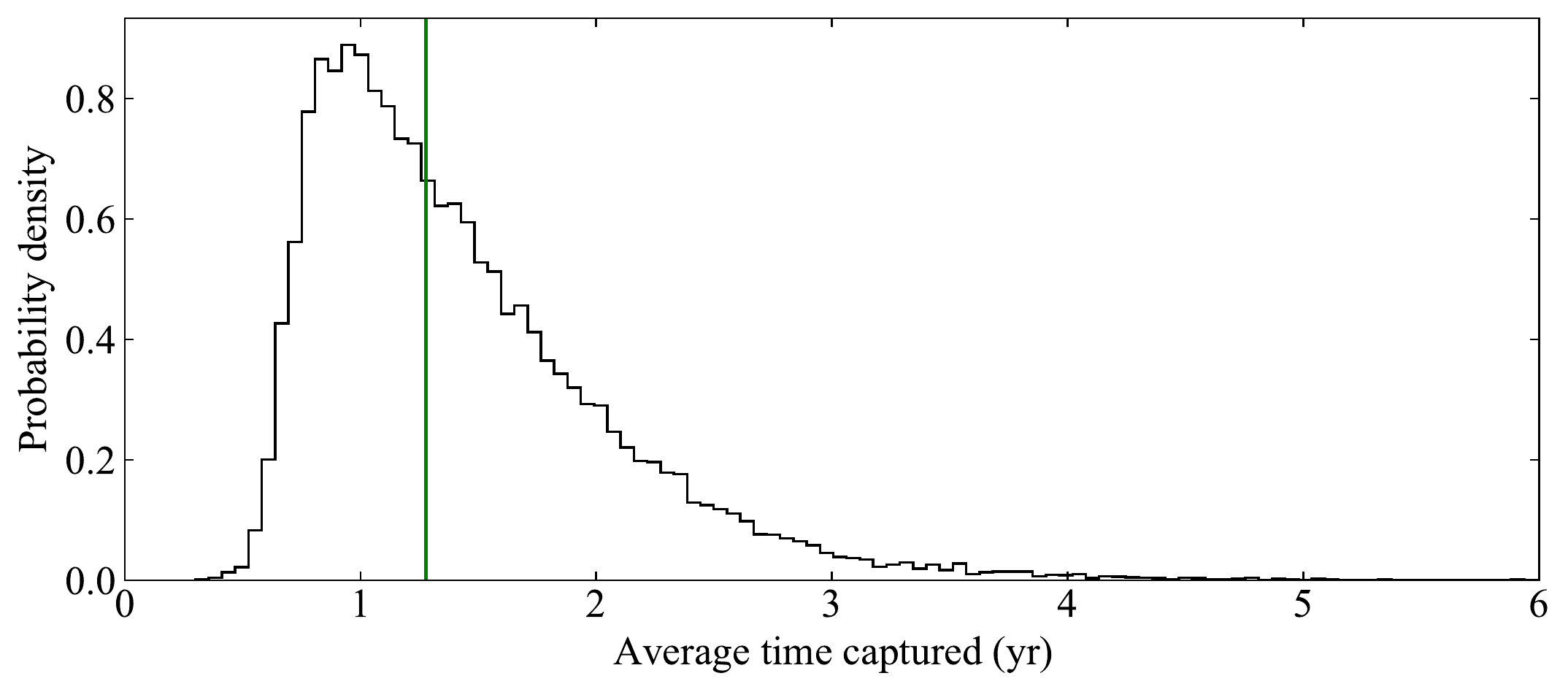}
         \includegraphics[width=\linewidth]{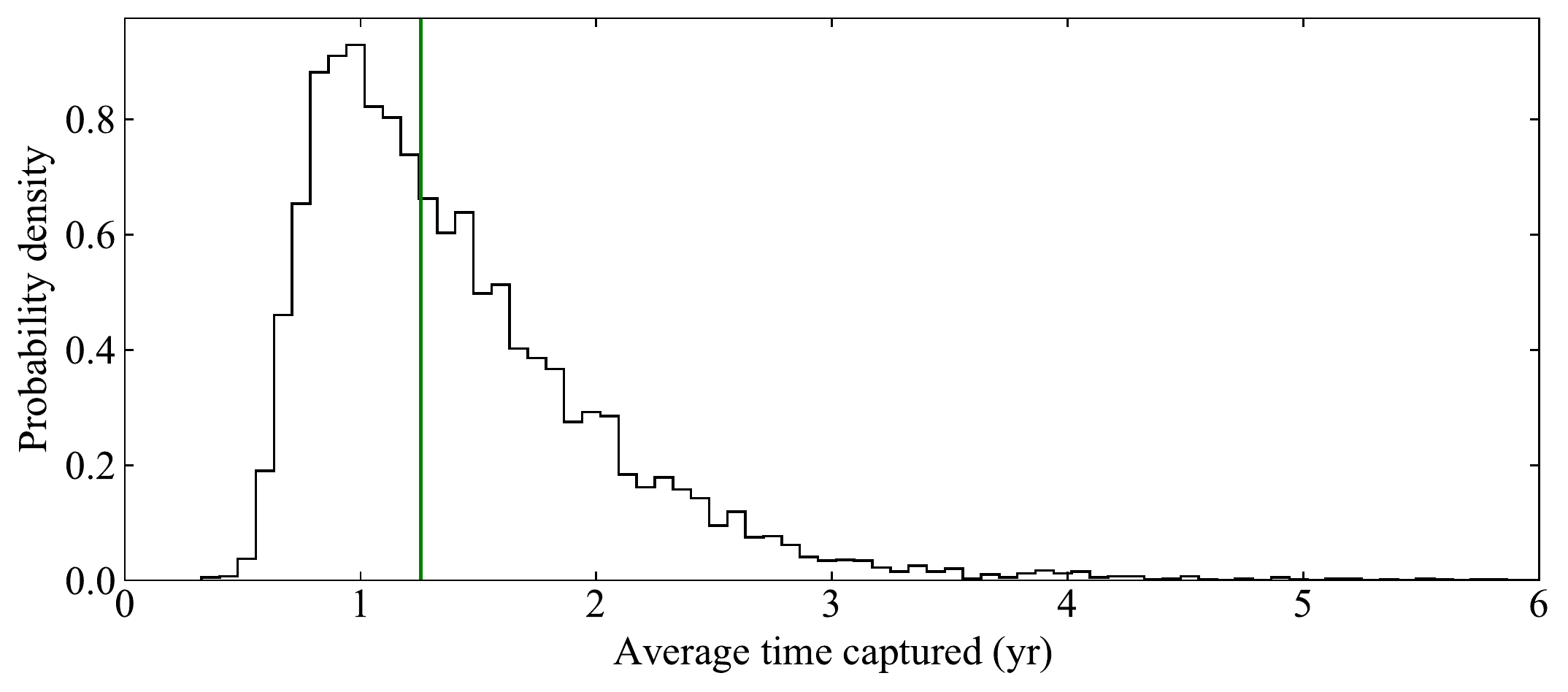}
         \includegraphics[width=\linewidth]{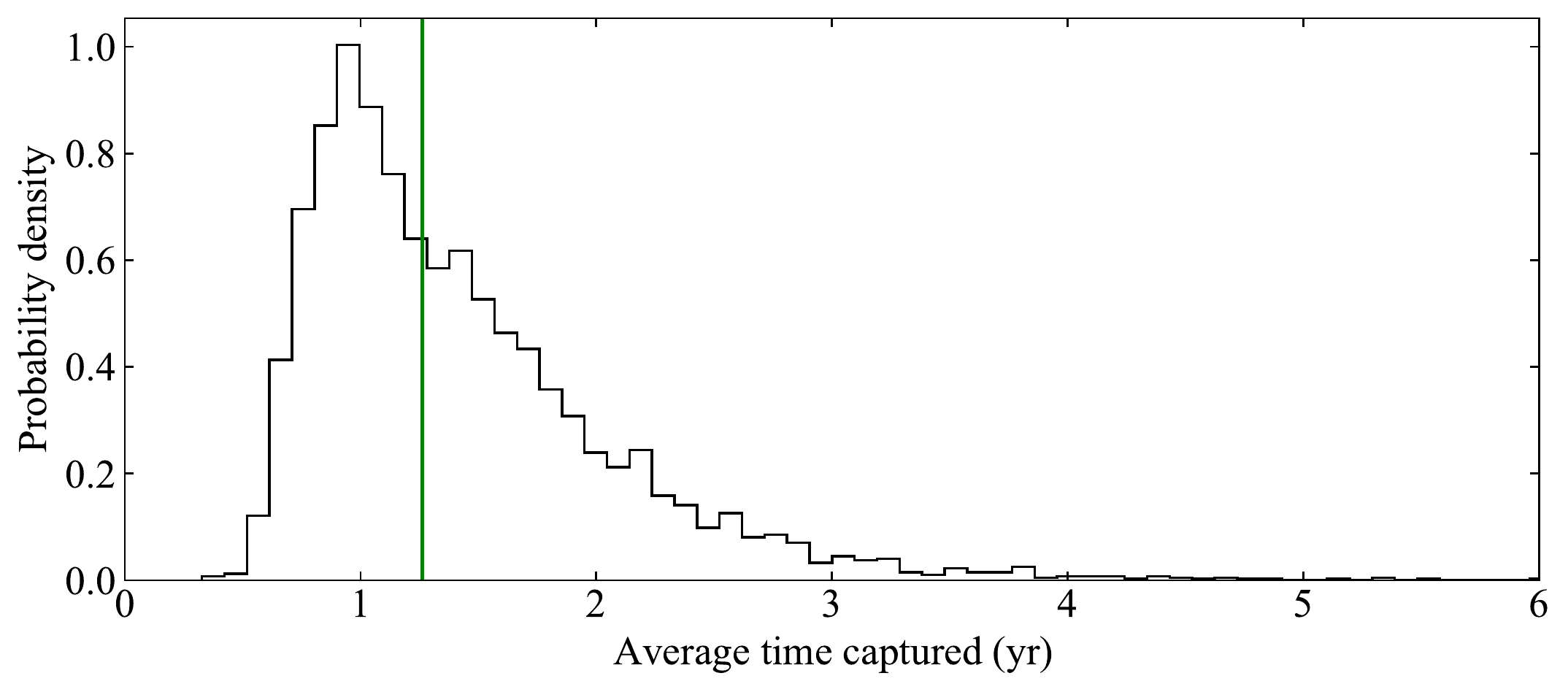}
         \caption{Distribution of average durations of temporary captures by Jupiter. {\em Upper panel:} For 2020~MK$_{4}$, 
                  29P/Schwassmann-Wachmann~1 ({\em second to upper panel}), P/2008~CL94 (Lemmon) ({\em second to bottom panel}), 
                  and  P/2010~TO20 (LINEAR-Grauer) ({\em bottom panel}). Median values are displayed as vertical green lines. The 
                  bins were computed using the Freedman and Diaconis rule implemented in NumPy \citep{2011CSE....13b..22V,
                  2020NumPy-Array}. In the histogram, we use counts to form a probability density so the area under the histogram 
                  will sum to one.
                 }
         \label{capdist}
      \end{figure}

         Using a sample of 20\,000 pairs of Gaussianly distributed control orbits based on the Cartesian vectors in 
         Tables~\ref{vector} and \ref{vector29P} and integrated backwards in time for 5\,000~yr (a similar sample integrated for 
         10\,000~yr produces nearly the same results), we have studied how the distribution is for minimum approach distances and 
         also the relative position of Jupiter during such close encounters. The top panel of Figure~\ref{ddist} shows the 
         distribution; the median and the 16th and 84th percentiles of the minimum approach distance distribution are 
         $0.27_{-0.15}^{+0.21}$~AU. Our calculations show that close encounters between these objects under one Lunar distance are 
         possible (the closest flyby was at about 363\,000~km), but the most surprising result is that the probability of this 
         pair experiencing a close encounter (following the distribution in Fig.~\ref{ddist}, top panel) while 2020~MK$_{4}$ has a 
         negative Jovicentric energy is 2.7\% and most of these temporary captures were within 100 Jovian radii. About 63\% of 
         close encounters take place within 1 Hill radius of Jupiter. The distribution of the durations of many of these capture 
         events is shown in Fig.~\ref{capdist} (top panel and second panel from the top). Comet 29P tends to experience slightly 
         longer captures than 2020~MK$_{4}$. 
%
%
%
%
%
%
%
%

         In general, (temporary or long-term) capture (by Jupiter) is a very low probability event. On the other hand, the fact 
         that close encounters between 2020~MK$_{4}$ and 29P may have taken place when one or both of these objects were temporary 
         satellites of Jupiter opens the door to an alternative dynamical scenario for the origin of 2020~MK$_{4}$, that of a 
         release by 29P with the gravitational assistance of Jupiter during a capture episode. Such a scenario has been observed 
         in the past, for example the tidal disruption suffered by comet Shoemaker-Levy~9 in 1992 (see for example 
         \citealt{1993IAUC.5800....1N}). Comet Shoemaker-Levy~9 was a Jovian satellite prior to impact \citep{1994PhDT........47B,
         1995Icar..118..155B}. With the available data, an origin for 2020~MK$_{4}$ during a tidal (or binary) disruption event 
         triggered by Jupiter on 29P cannot be excluded. Binary comets are rare, but they are known to exist 
         \citep{2017Natur.549..357A,2020A&A...643A.152A}; comet 288P/(300163) 2006 VW$_{139}$ could even be a triple 
         \citep{2020DPS....5221701K}.
         
         In order to gain a better understanding of the role of Jupiter on the evolution of the objects shown in 
         Table~\ref{elements}, we performed a similar study for the pairs P/2008~CL94 (Lemmon) and 2020~MK$_{4}$ as well as 
         P/2010~TO20 (LINEAR-Grauer) and 2020~MK$_{4}$. The middle and bottom panels of Figure~\ref{ddist} summarize our results 
         that use initial conditions from Tables~\ref{vectorCL94} and \ref{vectorTO20}. Close flybys when one or both members of 
         the pair were temporary satellites of Jupiter are found in both cases with respective probabilities of 0.023 and 0.026. 
         The distributions of the durations of the observed temporary capture events are displayed in Fig.~\ref{capdist}. For 
         comparison, the jovicentric orbital periods of known satellites of Jupiter range from slightly above 7~h (for Metis) to 
         2.17~yr (for S/2003~J~23).\footnote{\url{https://minorplanetcenter.net/mpec/K21/K21BD6.html}} 

         The topic of temporary capture of cometary objects by Jupiter has already been studied in the past (see for example 
         \citealt{1981A&A....94..226C}). Most known Jovian satellites have orbital periods close to 2~yr and move along very 
         elongated and inclined paths \citep{2003Natur.423..261S} so the objects discussed here may not complete one revolution 
         around Jupiter before returning to interplanetary space. Following \citet{2017Icar..285...83F}, we may consider these 
         episodes as linked to temporarily-captured flybys, not temporarily-captured orbiters. Although temporarily-captured 
         orbiter episodes in which the object completes one orbit around the planet are also possible.

         As for the possible past orbital evolution of 2020~MK$_{4}$ neglecting the possibility that it may have had an origin 
         within the 29P--P/2008~CL94--P/2010~TO20 cometary complex, the results of longer integrations using MCCM to generate 
         initial conditions indicate (see Fig.~\ref{origin}, left panel) that the probability of 2020~MK$_{4}$ having been 
         captured from interstellar space during the past 0.5~Myr could be 0.49$\pm$0.04. This result together with the previous 
         one for the probability of ejection indicate that the orbital evolution of 2020~MK$_{4}$ was as unstable in the past 
         as it will be in the future.

   \section{Discussion\label{discussion}}
      The origin of objects, such as 2020~MK$_{4}$, was thought to be in the trans-Neptunian or Edgeworth-Kuiper belt (see for 
      example \citealt{1980MNRAS.192..481F,1997Icar..127...13L}), but it is now generally accepted that this population may have 
      its source in the scattered belt (see for example \citealt{2009Icar..203..140D,2015A&A...573A.102B,2020CeMDA.132...36D,
      2021Icar..35814201R}); however, it is also important to consider the analysis in \citet{2018AJ....156..232G}. Although 
      cometary activity (either continuous or in the form of outbursts) has been detected at 30.7~AU for C/1995~O1 (Hale-Bopp) 
      \citep{2011A&A...531A..11S}, at 28.1~AU for 1P/Halley \citep{2004A&A...417.1159H}, and at 23.7~AU for C/2017~K2 (PANSTARRS) 
      \citep{2017ApJ...847L..19J}, so far no member of the trans-Neptunian belt (cold or scattered) has been recognized as active 
      (see for example \citealt{2019A&A...621A.102C}). In addition to comets, the only group of distant minor bodies that includes 
      known active objects is that of the centaurs. 

      According to JPL's SBDB search engine, as of March 5, 2021, there have been 547 known centaurs (objects with orbits between 
      Jupiter and Neptune, 5.5~AU $< a <$ 30.1~AU). The list of known active centaurs maintained by Y.~R. 
      Fern\'andez\footnote{\url{https://physics.ucf.edu/~yfernandez/cometlist.html\#ce}} includes 34 objects, not counting 
      (523676) 2013~UL$_{10}$ \citep{2018A&A...620A..93M} and the one discussed here, 2020~MK$_{4}$. Therefore, about 6.6\% of the
      known centaurs have been observed at some point losing mass and displaying a cometary physical appearance. Not counting 
      29P/Schwassmann-Wachmann~1, the first active centaur to be identified as such was 95P/Chiron \citep{1990Icar...83....1H}.
      Outbursts are sometimes associated with the ejection of sizable fragments, as in the case of 174P/Echeclus (see for 
      example \citealt{2008A&A...480..543R,2019AJ....158..255K}).  

      Figure~\ref{STevolution} shows that the objects in Table~\ref{elements} had a very chaotic dynamical past and that their 
      future orbital evolution will be equally chaotic, including very close encounters with Jupiter and perhaps also Saturn. In 
      addition, Fig.~\ref{ddist} indicates that these objects may have experienced relatively close mutual flybys and in some 
      cases such close encounters may have taken place when one or both of the objects involved were temporarily trapped by 
      Jupiter's gravitational pull, as shown in Fig.~\ref{capdist}. Therefore, Jupiter plays a central role in the dynamics of 
      this group of objects. 

\begin{figure}
        \centering
         \includegraphics[width=\linewidth]{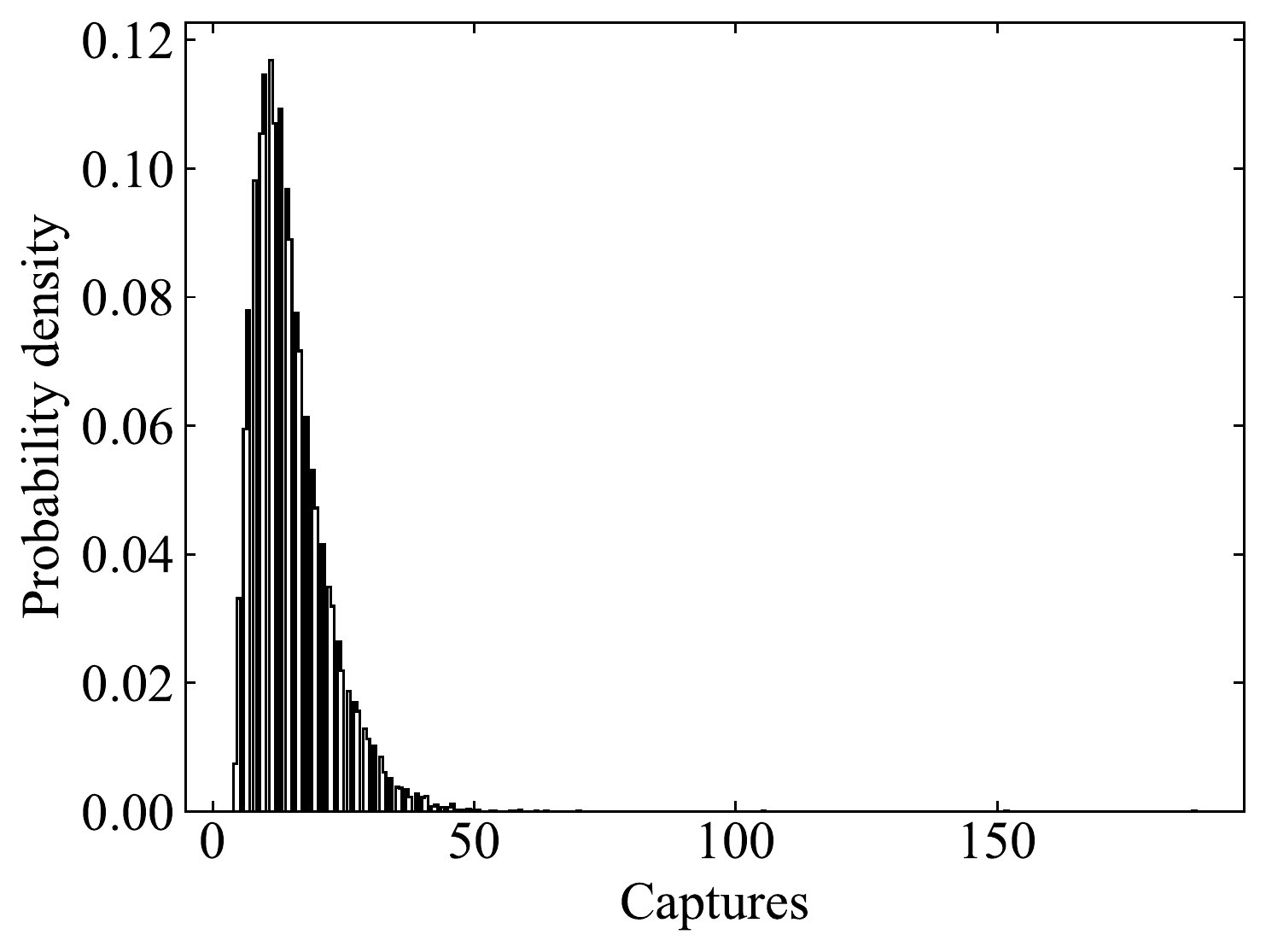}
         \includegraphics[width=\linewidth]{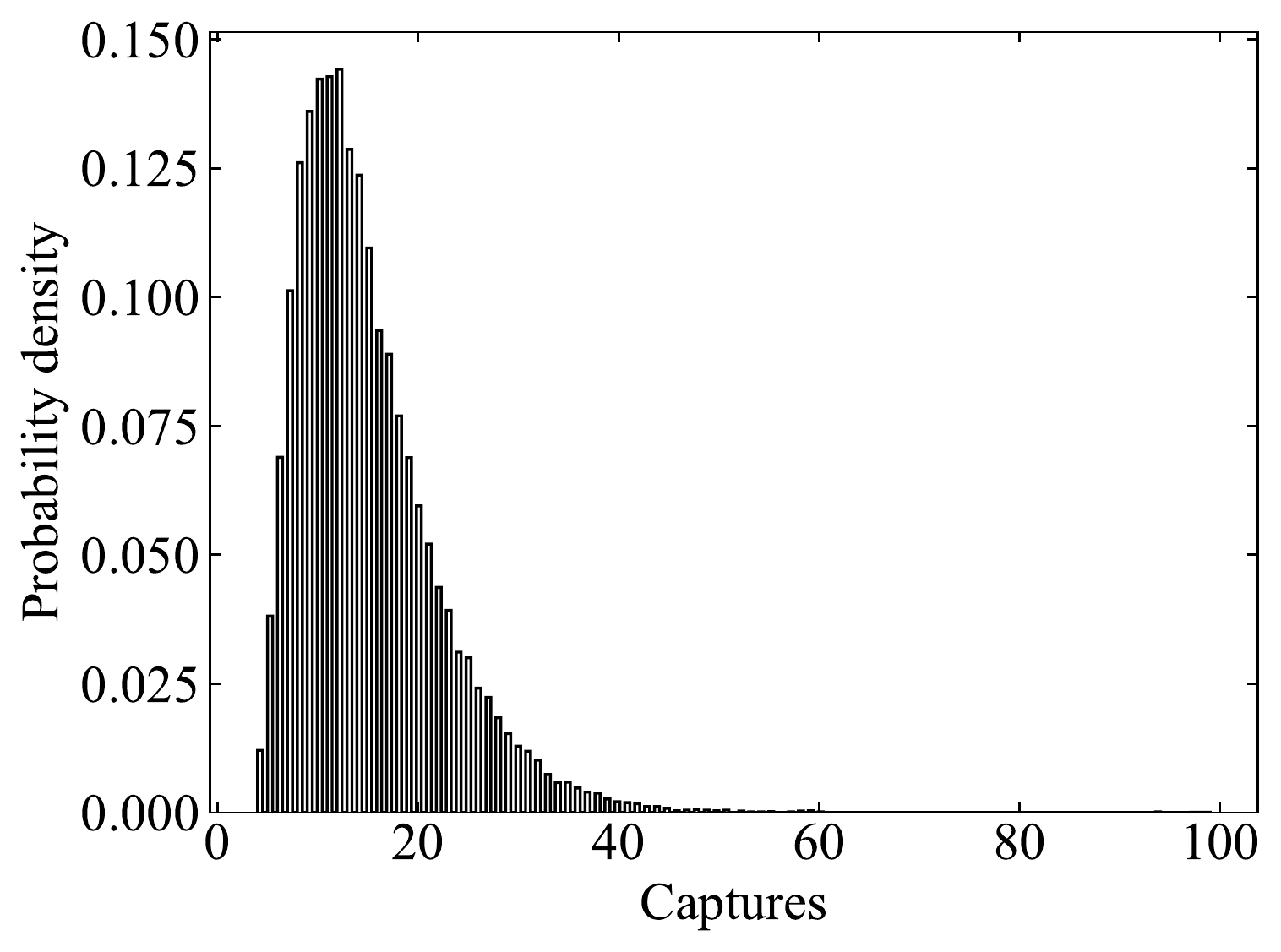}
         \caption{Distribution of the number of capture events by Jupiter per simulation. {\em Upper panel:} For 2020~MK$_{4}$ and 
                  for comet 29P/Schwassmann-Wachmann~1 ({\em bottom panel}). The bins were computed using the Freedman and 
                  Diaconis rule implemented in NumPy \citep{2011CSE....13b..22V,2020NumPy-Array}. In the histogram, we use counts 
                  to form a probability density so the area under the histogram will sum to one.
                 }
         \label{capture}
      \end{figure}

      In general, our results indicate that the evolution of these minor bodies can only be reliably predicted a few thousand 
      years into the past or the future, which is consistent with the conclusions in the extensive study of 
      \citet{2021Icar..35814201R}. Although \citet{2021Icar..35814201R} argue that mean-motion resonances with the giant planets 
      may stabilize some of the orbits in this region and they provide some examples, for the set of objects studied here, 
      resonant behavior fails to appear or at least its strength is not enough to mitigate their chaotic orbital evolution. The 
      dynamical behavior observed is compatible with that of the centaurs experiencing generalized diffusion as discussed by 
      \citet{2009Icar..203..155B}. In addition, relatively close and recurrent encounters between these objects are possible, and 
      they are not impeded by resonances. Temporary captures by Jupiter of these objects are also recurrent (see 
      Fig.~\ref{capture}). 
%
%
%
%

      The orbital context analysis presented in Sect.~4.1 opens the possibility to close encounters between objects of this 
      dynamical class as the mutual nodal distances are small. This theoretical possibility is confirmed by our $N$-body 
      simulations, which indicate that close approaches at distances on the order of 10$^{5}$~km (or 10$^{-3}$~AU) are indeed 
      possible.\ However, this is the typical size of the coma of comet 29P when in outburst (see for example 
      \citealt{2008A&A...485..599T}) and this has the potential for a physical interaction between material in the comae of 
      these objects during close encounters. This issue has never been considered before in the literature and it may accelerate 
      the erosion rate of a cometary object as one active object may periodically penetrate the nebulous envelope of another and 
      both experience mutual enhanced surface bombardment episodes. 

      Although our dynamical results cannot determine the origin of any of the objects studied in the scattered belt, they uncover an 
      alternative or perhaps complementary scenario that may lead to increasing the population of objects following 29P-like 
      orbits, that of the in situ production of these objects thanks to interactions with Jupiter of some precursor bodies. The 
      existence of pairs of objects with small values of their mutual nodal distances and correlated orbits discussed in Sect.~4.1 
      provides additional support to this scenario when considering the very short dynamical lifetimes of these bodies.

      As for the nature of the cometary-like activity of 2020~MK$_{4}$ presented in this paper, it seems to be irregular, not
      continuous. The data\footnote{\url{https://www.minorplanetcenter.net/db_search/show_object?utf8=\%E2\%9C\%93&object_id=2020+MK4}} 
      available from the Minor Planet Center (MPC, \citealt{2016IAUS..318..265R,2019AAS...23324503H})\footnote{\url{https://minorplanetcenter.net}} 
      indicate that its apparent magnitude went from $G=21.12$~mag on June 15, 2020, to $G=18.1$~mag on July 10, 2020, returning 
      to $G=22.5$~mag by November 9, 2020. The data from the MPC suggest that the outburst may have stopped at some time between 
      early September and November. Irregular activity is often found among centaurs \citep{2009AJ....137.4296J}. Follow-up 
      observations of this centaur are necessary to understand the nature of its activity.

   \section{Conclusions\label{conclusions}}
      In this paper, we have presented observations of 2020~MK$_{4}$ obtained with JKT and IAC80 which we have used to establish
      the active status of this centaur and to derive its colors. Its current orbital context has been outlined and its past, 
      present, and future orbital evolution has been explored using direct $N$-body simulations. The object was originally 
      selected to carry out this study because its orbit determination resembles that of comet 29P/Schwassmann-Wachmann~1 (see 
      Table~\ref{elements}) and its first published observations hinted at an ongoing outburst event. Our conclusions can be 
      summarized as follows.
      \begin{enumerate}
         \item We show that the PSF of 2020~MK$_{4}$ is nonstellar and this confirms the presence of cometary-like activity in the 
               form of a conspicuous coma.
         \item Centaur 2020~MK$_{4}$ is neutral-gray in color. The values of its color indexes, $(g'-r')=0.42\pm0.04$ and 
               $(r'-i')=0.17\pm0.04$, are similar to the solar ones. These values are typical for active centaurs.  
         \item A lower limit for the absolute magnitude of the nucleus of 2020~MK$_{4}$ is $H_{g}=11.30\pm0.03$~mag which, for an 
               albedo in the range of 0.1--0.04, gives an upper limit for its size in the interval (23,~37)~km.
         \item The orbital evolution of 2020~MK$_{4}$ is very chaotic and it may eventually be ejected from the Solar System. It
               had a very chaotic dynamical past as well.
         \item Both 2020~MK$_{4}$ and 29P may have been recurrent transient Jovian satellites. This also applies to P/2008~CL94 
               (Lemmon) and P/2010~TO20 (LINEAR-Grauer).
         \item Although the past, present, and future dynamical evolution of 2020~MK$_{4}$ resembles that of 29P, a more robust 
               orbit determination is needed to confirm or reject a possible relation between the two objects.
         \item Our analyses suggest that active minor bodies may experience close encounters in which they traverse each other's
               comae, experimenting mutual enhanced surface bombardment which may accelerate the erosion rate of these objects.
               Our analyses confirm that penetrating encounters in which 2020~MK$_{4}$ may travel across the coma of comet 29P are 
               possible. 
      \end{enumerate}    
      In summary, and based on the analysis of visible CCD images of 2020~MK$_{4}$, we confirm the presence of a coma of material 
      around a central nucleus. Its surface colors place this centaur among the most extreme members of the gray group. Although 
      the past, present, and future dynamical evolution of 2020~MK$_{4}$ resembles that of 29P, more data are required to confirm
      or reject a possible connection between the two objects and perhaps others.
      
   \begin{acknowledgements}
      We thank the referee for her/his constructive and detailed report that included very helpful suggestions regarding the
      presentation of this paper and the interpretation of our results. CdlFM and RdlFM thank S.~J. Aarseth for providing one of 
      the codes used in this research, A.~I. G\'omez de Castro for providing access to computing facilities, and S. Deen for 
      extensive comments on the existence of precovery images of 2020~MK$_{4}$. Part of the calculations and the data analysis 
      were completed on the Brigit HPC server of the `Universidad Complutense de Madrid' (UCM), and we thank S. Cano Als\'ua for 
      his help during this stage. This work was partially supported by the Spanish `Ministerio de Econom\'{\i}a y Competitividad' 
      (MINECO) under grant ESP2017-87813-R. JdL acknowledges support from MINECO under the 2015 Severo Ochoa Program SEV-2015-0548. 
      This article is based on observations made with the IAC80 telescope operated on the island of Tenerife by the Instituto de 
      Astrof\'{\i}sica de Canarias in the Spanish Observatorio del Teide and on observations obtained with the 1m JKT telescope 
      operated by the Southeastern Association for Research in Astronomy (saraobservatory.org) in the Spanish Observatorio del 
      Roque de los Muchachos on the island of La Palma. In preparation of this paper, we made use of the NASA Astrophysics Data 
      System, the ASTRO-PH e-print server, and the MPC data server.
   \end{acknowledgements}

   \bibliographystyle{aa} 
   \bibliography{refs2020MK4.bib}

   \begin{appendix}
      \section{Mutual nodal distances and uncertainty estimates\label{29Plikeelements}}
         The mutual nodal distance between two Keplerian trajectories with a common focus can be written as follows (see eqs.~16 and 17 in
         \citealt{2017CeMDA.129..329S}):
         \begin{equation}
            {\Delta}_{\pm} =  \frac{a_{2} \ (1 - e_{2}^2)}{1 \pm e_{2} \ \cos{\varpi_{2}}} - \frac{a_{1} \ (1 - e_{1}^2)}{1 \pm e_{1}
                                    \ \cos{\varpi_{1}}} \,,
            \label{noddis}
         \end{equation}
         where for prograde orbits the "+" sign refers to the ascending node and the "$-$" sign to the descending one, and
         \begin{equation}
            \resizebox{0.9\hsize}{!}{$
            \cos{\varpi_{1}} =  \frac{-\cos{\omega_{1}} \ (\sin{i_{1}}\ \cos{i_{2}} - \cos{i_{1}}\ \sin{i_{2}}\ \cos{\Delta\Omega}) +
                                      \sin{\omega_{1}}\ \sin{i_{2}}\ \sin{\Delta\Omega}}
                                     {\sqrt{1 - (\cos{i_{2}}\ \cos{i_{1}} + \sin{i_{2}}\ \sin{i_{1}}\ \cos{\Delta\Omega})^2}}
                                     $}
            \label{cosvarpi1}
         \end{equation}
         and
         \begin{equation}
            \resizebox{0.9\hsize}{!}{$
            \cos{\varpi_{2}} =  \frac{\cos{\omega_{2}} \ (\sin{i_{2}}\ \cos{i_{1}} - \cos{i_{2}}\ \sin{i_{1}}\ \cos{\Delta\Omega}) +
                                      \sin{\omega_{2}}\ \sin{i_{1}}\ \sin{\Delta\Omega}}
                                     {\sqrt{1 - (\cos{i_{2}}\ \cos{i_{1}} + \sin{i_{2}}\ \sin{i_{1}}\ \cos{\Delta\Omega})^2}}
                                     \,,$}
           \label{cosvarpi2}
         \end{equation}
         with $\Delta\Omega = \Omega_{2} - \Omega_{1}$, and $a_j$, $e_j$, $i_j$, $\Omega_j$, and $\omega_j$ ($j=1, 2$) are the 
         orbital elements of the orbits involved. In order to obtain the actual distributions of ${\Delta}_{\pm}$, we generated  
         sets of orbital elements for the virtual objects using data from JPL's SBDB. For example, the value of the semimajor 
         axis of a virtual object was computed using the expression $a_{\rm v}=a + \sigma_{a}\,r_{\rm i}$, where $a$ is the
         semimajor axis, $\sigma_{a}$ is the standard deviation, and $r_{\rm i}$ is a (pseudo) random number with a normal 
         distribution computed using NumPy \citep{2011CSE....13b..22V,2020NumPy-Array}. In order to calculate statistically 
         relevant values of ${\Delta}_{\pm}$, we computed median and 16th and 84th percentiles from a set of 10$^4$ pairs of 
         virtual objects for each pair.

      \section{Angular distances between pairs of orbital poles and perihelia\label{spaceorientations}}
         In order to understand the context of the orientations in space of the 29P-like orbits, we study the line of apsides of 
         their paths and the projection of their orbital poles onto the plane of the sky. In heliocentric ecliptic coordinates, 
         the longitude and latitude of an object at perihelion, $(l_q, b_q)$, are given by the following expressions: 
         $\tan{(l_q-\Omega)}=\tan\omega\,\cos{i}$ and $\sin{b_q}=\sin\omega\,\sin{i}$ (see for example 
         \citealt{1999ssd..book.....M}). On the other hand, the ecliptic coordinates of the pole are $(l_{\rm p}, b_{\rm p}) = 
         (\Omega-90\degr, 90\degr-i)$. The angular distances between pairs of orbital poles and perihelia (see 
         Fig.~\ref{orbinspace}) are given by the angles $\alpha_{q}$ and $\alpha_{\rm p}$:
         \begin{equation} 
            \cos{\alpha_{q}} = \cos{b_{q2}} \ \cos{b_{q1}} \ \cos{(l_{q2} - l_{q1})} + \sin{b_{q2}} \ \sin{b_{q1}} 
         \end{equation}
         and
         \begin{equation}
            \cos{\alpha_{\rm p}} = \cos{b_{\rm p2}} \ \cos{b_{\rm p1}} \ \cos{(l_{\rm p2} - l_{\rm p1})} + \sin{b_{\rm p2}} 
                                    \ \sin{b_{\rm p1}} \,,
         \end{equation}
         where the sets of orbital elements for the virtual objects were generated and the uncertainties were computed 
         as described above.

      \section{Input data\label{Adata}}
         Here, we include the barycentric Cartesian state vectors of the four objects in Table~\ref{elements}. These vectors and 
         their uncertainties were used to carry out the calculations discussed above and to generate the figures that display 
         the time evolution of the various orbital parameters and the histograms of the close encounters of pairs of objects. For 
         example, a new value for the $X$-component of the state vector was computed as $X_{\rm c} = X + \sigma_X \ r$, where $r$ is 
         an univariate Gaussian random number, and $X$ and $\sigma_X$ are the mean value and its 1$\sigma$ uncertainty in the 
         corresponding table. 
%
%
     \begin{table}
      \centering
      \fontsize{8}{12pt}\selectfont
      \tabcolsep 0.15truecm
      \caption{\label{vector}Barycentric Cartesian state vector of 2020~MK$_{4}$: Components and associated 1$\sigma$
               uncertainties.
              }
      \begin{tabular}{ccc}
       \hline
        Component                         &   &    value$\pm$1$\sigma$ uncertainty                                \\
       \hline
        $X$ (AU)                          & = &    2.476823027013801$\times10^{+0}$$\pm$2.63888043$\times10^{-5}$ \\
        $Y$ (AU)                          & = & $-$5.649556533725540$\times10^{+0}$$\pm$5.00052909$\times10^{-5}$ \\
        $Z$ (AU)                          & = & $-$6.724625273177226$\times10^{-1}$$\pm$6.68127413$\times10^{-6}$ \\
        $V_X$ (AU/d)                      & = &    6.344171798287663$\times10^{-3}$$\pm$5.52645958$\times10^{-7}$ \\
        $V_Y$ (AU/d)                      & = &    2.598886274097145$\times10^{-3}$$\pm$1.15080359$\times10^{-6}$ \\
        $V_Z$ (AU/d)                      & = &    2.738174058059657$\times10^{-4}$$\pm$1.27180464$\times10^{-7}$ \\
       \hline
      \end{tabular}
      \tablefoot{Data are referred to as epoch 2459000.5, 31-May-2020 00:00:00.0 TDB (J2000.0 ecliptic and equinox). 
                 Source: JPL's SBDB.
                }
     \end{table}
%
%

%
%
     \begin{table}
      \centering
      \fontsize{8}{12pt}\selectfont
      \tabcolsep 0.15truecm
      \caption{\label{vector29P}Barycentric Cartesian state vector of comet 29P/Schwassmann-Wachmann~1: Components and associated 
               1$\sigma$ uncertainties.
              }
      \begin{tabular}{ccc}
       \hline
        Component                         &   &    value$\pm$1$\sigma$ uncertainty                                 \\
       \hline
        $X$ (AU)                          & = &    4.820127849857867$\times10^{+0}$$\pm$2.26195239$\times10^{-7}$  \\
        $Y$ (AU)                          & = &    3.096547270522562$\times10^{+0}$$\pm$2.26291332$\times10^{-7}$  \\
        $Z$ (AU)                          & = &    9.315292658149408$\times10^{-1}$$\pm$1.80747737$\times10^{-7}$  \\
        $V_X$ (AU/d)                      & = & $-$3.818139825384470$\times10^{-3}$$\pm$2.98663624$\times10^{-10}$ \\
        $V_Y$ (AU/d)                      & = &    6.187940318090741$\times10^{-3}$$\pm$1.63227588$\times10^{-10}$ \\
        $V_Z$ (AU/d)                      & = &    2.243304452195744$\times10^{-4}$$\pm$2.29798018$\times10^{-10}$ \\
       \hline
      \end{tabular}
      \tablefoot{Data are referred to as epoch 2459000.5, 31-May-2020 00:00:00.0 TDB (J2000.0 ecliptic and equinox).
                 Source: JPL's SBDB.
                }
     \end{table}
%
%

%
%
     \begin{table}
      \centering
      \fontsize{8}{12pt}\selectfont
      \tabcolsep 0.15truecm
      \caption{\label{vectorCL94}Barycentric Cartesian state vector of comet P/2008 CL94 (Lemmon): Components and associated 
               1$\sigma$ uncertainties.
              }
      \begin{tabular}{ccc}
       \hline
        Component                         &   &    value$\pm$1$\sigma$ uncertainty                                \\
       \hline
        $X$ (AU)                          & = &    1.459836826464546$\times10^{+0}$$\pm$2.48312839$\times10^{-3}$ \\
        $Y$ (AU)                          & = &    5.336008875429940$\times10^{+0}$$\pm$8.60065292$\times10^{-4}$ \\
        $Z$ (AU)                          & = &    5.352334095577819$\times10^{-1}$$\pm$3.08664106$\times10^{-4}$ \\
        $V_X$ (AU/d)                      & = & $-$7.484540791199146$\times10^{-3}$$\pm$4.84686920$\times10^{-7}$ \\
        $V_Y$ (AU/d)                      & = &    1.423074128372227$\times10^{-3}$$\pm$3.12580298$\times10^{-6}$ \\
        $V_Z$ (AU/d)                      & = &    7.781756239073725$\times10^{-4}$$\pm$3.54702110$\times10^{-7}$ \\
       \hline
      \end{tabular}
      \tablefoot{Data are referred to as epoch 2459000.5, 31-May-2020 00:00:00.0 TDB (J2000.0 ecliptic and equinox). 
                 Source: JPL's SBDB.
                }
     \end{table}
%
%

%
%
     \begin{table}
      \centering
      \fontsize{8}{12pt}\selectfont
      \tabcolsep 0.15truecm
      \caption{\label{vectorTO20}Barycentric Cartesian state vector of comet P/2010 TO20 (LINEAR-Grauer): Components and 
               associated 1$\sigma$ uncertainties.
              }
      \begin{tabular}{ccc}
       \hline
        Component                         &   &    value$\pm$1$\sigma$ uncertainty                                \\
       \hline
        $X$ (AU)                          & = & $-$2.760378798588101$\times10^{+0}$$\pm$1.87330516$\times10^{-3}$ \\
        $Y$ (AU)                          & = & $-$4.981877686502737$\times10^{+0}$$\pm$5.04623641$\times10^{-4}$ \\
        $Z$ (AU)                          & = & $-$7.533566693101323$\times10^{-2}$$\pm$7.13336260$\times10^{-5}$ \\
        $V_X$ (AU/d)                      & = &    6.530991396932470$\times10^{-3}$$\pm$1.34995329$\times10^{-6}$ \\
        $V_Y$ (AU/d)                      & = & $-$3.198756786861413$\times10^{-3}$$\pm$1.72833828$\times10^{-6}$ \\
        $V_Z$ (AU/d)                      & = & $-$2.946577560510802$\times10^{-4}$$\pm$2.25495604$\times10^{-8}$ \\
       \hline
      \end{tabular}
      \tablefoot{Data are referred to as epoch 2459000.5, 31-May-2020 00:00:00.0 TDB (J2000.0 ecliptic and equinox). 
                 Source: JPL's SBDB.
                }
     \end{table}
%
%

   \end{appendix}

\end{document}